\documentclass[fleqn,usenatbib,useAMS]{mnras}

\usepackage[T1]{fontenc}

\DeclareRobustCommand{\VAN}[3]{#2}
\let\VANthebibliography\thebibliography
\def\thebibliography{\DeclareRobustCommand{\VAN}[3]{##3}\VANthebibliography}

\usepackage{hyperref}
\usepackage{subcaption}
\usepackage{graphicx}	
\usepackage{amsmath}	
\usepackage{amssymb}	
\usepackage{tablefootnote}
\usepackage{gensymb}
\usepackage{lineno}

\def\mass{$M_{\rm{\odot}}$}
\def\kms{km~s$^{-1}$}

\def\nifs{$^{56}$Ni}

\def\phn{\phantom{0}}

\usepackage{newtxtext,newtxmath}

\title[Intrinsic rate of SNe~Ia excesses]{The detection efficiency of type Ia supernovae from the Zwicky Transient Facility: Limits on the intrinsic rate of early flux excesses }

\author[M. R. Magee et al.]{
M. R. Magee$^{1,2}$\thanks{E-mail: mrmagee.astro@gmail.com},
C. Cuddy$^{1}$,
K. Maguire$^{1}$,
M. Deckers$^{1}$,
S. Dhawan$^{3}$,
C. Frohmaier$^{4}$,
A. A. Miller$^{5}$,\newauthor
J. Nordin$^{6}$,
M. W. Coughlin${^7}$,
F. Feinstein$^{8}$,
R. Riddle$^{9}$
\\
$^{1}$School of Physics, Trinity College Dublin, The University of Dublin, Dublin 2, Ireland \\ 
$^{2}$Institute of Cosmology and Gravitation, University of Portsmouth, Burnaby Road, Portsmouth, PO1 3FX, UK \\
$^{3}$The Oskar Klein Centre for Cosmoparticle Physics, Department of Physics, Stockholm University, SE-10691 Stockholm, Sweden\\
${^4}$School of Physics and Astronomy, University of Southampton, Southampton, SO17 1BJ, UK\\
$^{5}$Center for Interdisciplinary Exploration and Research in Astrophysics and Department of Physics and Astronomy, Northwestern
University, \\1800 Sherman Ave, Evanston, IL 60201, USA\\
$^{6}$Institute of Physics, Humboldt-Universitat zu Berlin, Newton str. 15, 12489 Berlin, Germany\\
$^{7}$School of Physics and Astronomy, University of Minnesota, Minneapolis, Minnesota 55455, USA\\
$^{8}$Aix Marseille Univ, CNRS/IN2P3, CPPM, Marseille, France\\
$^{9}$Caltech Optical Observatories, California Institute of Technology, Pasadena, CA 91125, USA\\
}

\date{Accepted 2022 April 09. Received 2022 April 09; in original form 2021 November 25.}

\pubyear{2022}

\begin{document}
\label{firstpage}
\pagerange{\pageref{firstpage}--\pageref{lastpage}}
\maketitle

\begin{abstract}
Samples of young type Ia supernovae have shown `early excess' emission in a few cases. Similar excesses are predicted by some explosion and progenitor scenarios and hence can provide important clues regarding the origin of thermonuclear supernovae. They are however, only predicted to last up to the first few days following explosion. It is therefore unclear whether such scenarios are intrinsically rare or if the relatively small sample size simply reflects the difficulty in obtaining sufficiently early detections. To that end, we perform toy simulations covering a range of survey depths and cadences, and investigate the efficiency with which young type Ia supernovae are recovered. As input for our simulations, we use models that broadly cover the range of predicted luminosities. Based on our simulations, we find that in a typical three day cadence survey, only $\sim$10\% of type Ia supernovae would be detected early enough to rule out the presence of an excess. A two day cadence however, should see this increase to $\sim$15\%. We find comparable results from more detailed simulations of the Zwicky Transient Facility surveys. Using the recovery efficiencies from these detailed simulations, we investigate the number of young type Ia supernovae expected to be discovered assuming some fraction of the population come from scenarios producing an excess at early times. Comparing the results of our simulations to observations, we find the intrinsic fraction of type Ia supernovae with early flux excesses is $\sim28^{+13}_{-11}\%$.
\end{abstract}




\section{Introduction}
\label{sect:intro}

The optical sky is being studied in unprecedented detail thanks to modern all-sky surveys such as the All-Sky Automated Survey for Supernovae (ASAS-SN; \citealt{asassn}), Asteroid Terrestrial-impact Last Alert System (ATLAS; \citealt{atlas}), and Zwicky Transient Facility (ZTF; \citealt{ztf, graham--19, masci--19, dekany--20}). The increased cadence and area of these surveys has allowed for new and unique behaviours to be observed, including for objects that were once thought to be relatively homogeneous, such as type Ia supernovae (SNe~Ia). Consensus in the literature points to thermonuclear explosions of white dwarfs in binary systems as the origin of SNe~Ia \citep{livio--18, wang--18, jha--19, soker--19}. There is little consensus however, over exactly the nature of the progenitor and explosion scenarios. In addition, observations at increasingly early times are highlighting new and poorly understood phenomena and therefore could provide key diagnostic evidence for discriminating between different scenarios (e.g. \citealt{11fe--nature, jiang--2017}).

\par

The first in-depth investigation of the progenitor system's impact on the early light curves of SNe~Ia was provided by \cite{kasen--10}, who studied the emission generated following interaction between the SN ejecta and companion star. During the interaction, the ejecta becomes shock-heated and eventually results in UV and optical emission lasting up to a few days as the shocked ejecta expands and cools. The strength of this emission is determined by the properties of the binary system and viewing angle of the observer. Following from the arguments presented by \cite{kasen--10}, \cite{liu--2015c} investigate the range of companion interaction signatures that may be expected from binary population synthesis calculations. Related to the companion interaction scenario, \cite{piro-16} investigate interaction between the SN ejecta and circumstellar material. This scenario behaves in a similar manner to that of companion interaction and therefore produces similar observational signatures, namely an excess of UV and optical flux within the first few days of explosion. 

\par

Instead of interaction, an excess of flux at early times can also be produced by certain explosion scenarios. The double detonation scenario involves the accretion of a helium shell on to the surface of the white dwarf \citep{livne--90, livne--91, woosley--94}. During the explosion, burning within the helium shell can lead to the production of short-lived radioactive isotopes, such as \nifs{}, $^{52}$Fe, and $^{48}$Cr. The decay of these isotopes in the outer ejecta drives the early emission for the first few days following explosion \citep{noebauer-17, jiang--2017, polin--19, magee--21}. Depending on the mass and composition of the helium shell, a wide variety of behaviours at early times can be produced \citep{maeda--18, magee--21}. Similarly, shells or clumps of \nifs{} present in the outer ejecta could also lead to excess emission at early times, before the primary \nifs{} mass in the inner ejecta powers the main rise of the light curve \citep{dimitriadis--19, magee--20b}. Such clumps could arise from deflagrations in Chandrasekhar-mass white dwarfs \citep{seitenzahl--16}

\par

Observations of SNe~Ia within these crucial early phases are becoming increasingly routine and, as such, dedicated searches for the signatures of these scenarios have been undertaken. \cite{cao--15} present observations of iPTF14atg, the first SN~Ia discovered with a prominent UV flash at early times. The origin of this early excess is contentious, with arguments presented for and against interaction scenarios \citep{cao--15, liu--2015c, kromer--16}. Subsequent in-depth analyses for the few SNe~Ia with clear early excesses have demonstrated the wide variety of early emission behaviours observed \citep{jiang--2017, hosseinzadeh--17, de--19, dimitriadis--19, shappee--2019, miller--19yvq, ni-22}. It is likely that more than one scenario is required to explain the full diversity observed \citep{magee--21}.

\par

While a handful of SNe~Ia with early excesses are now known, it is not clear whether the scenarios producing such excesses are intrinsically rare or if we simply lack sufficient coverage at early times to detect them. In this work, we perform a direct investigation of the efficiency with which early exccesses in the light curves of SNe~Ia are recovered. We further use this analysis to place limits on the rate of objects with such excesses. In Sect.~\ref{sect:model_lcs} we discuss the different model scenarios used as part of this work. Section~\ref{sect:simsurvey} provides a description of \texttt{simsurvey}\footnote{https://simsurvey.readthedocs.io/en/latest/} \citep{simsurvey}, which is used to perform our simulations of surveys with various different observing strategies. In Sect.~\ref{sect:toy_sims} we present the results of a series of toy simulations that directly explore the impact of cadence and observing depth on the recovery efficiency of young SNe~Ia. Section~\ref{sect:ztf_sims} presents the results of simulations based on observing logs from ZTF. Section~\ref{sect:rates} uses our simulations of the actual ZTF observing strategy to calculate the rate of SNe~Ia from ZTF and places limits on the intrinsic fraction of SNe~Ia that may produce early flux excesses. We discuss our results in Sect.~\ref{sect:discuss}. Finally, in Sect.~\ref{sect:conclusions}, we present our conclusions.

%

\section{Model SN~Ia light curves}
\label{sect:model_lcs}

Here, we outline the different models used as part of our survey simulations. These models were chosen due to the great diversity in their early time behaviours and general similarities to observed flux excesses, which will allow us to place meaningful constraints on the expected rates of early time flux excesses in SNe~Ia. 

\par

\subsection{Model scenarios}

\begin{figure}
\centering
\includegraphics[width=\columnwidth]{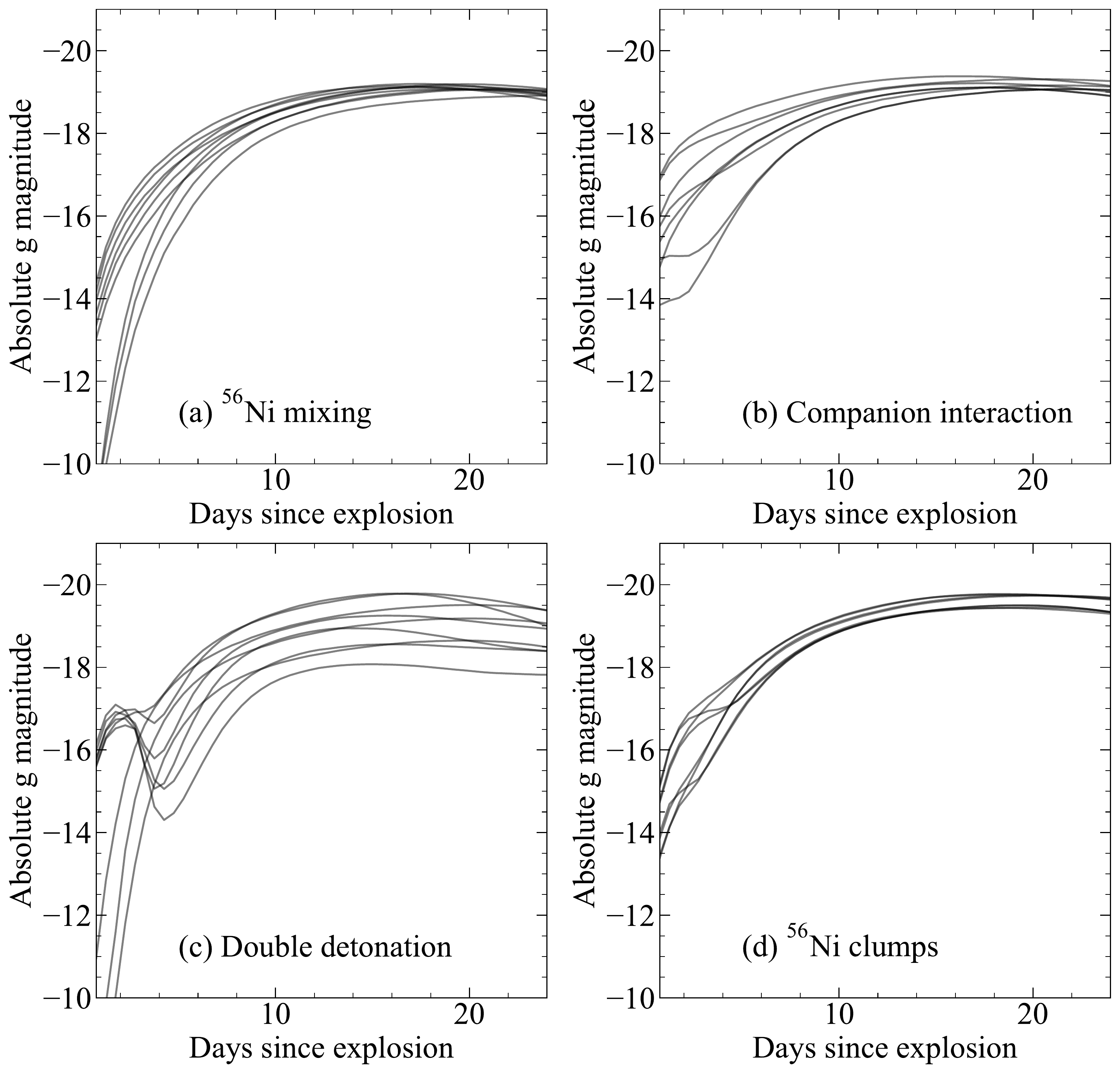}
\caption{Representative examples of light curves for the four model scenarios considered in this work.
}
\label{fig:scenario_examples}
\centering
\end{figure}

Each model is based on one of the following four scenarios: \nifs{} mixing, companion interaction, double detonation, and \nifs{} clumps. The free parameters for each scenario are given in Table~\ref{tab:models}. Here we provide a brief overview of each scenario. References for more detailed information are given in Table~\ref{tab:models}. A selection of representative light curves are also shown in Fig.~\ref{fig:scenario_examples}.

\par

All model light curves were calculated using \textsc{turtls} \citep{magee--18, magee--20b, magee--21}. Briefly, \textsc{turtls} is a one-dimensional Monte-Carlo radiative transfer code for modelling thermonuclear SNe. The structure of the ejecta is a free parameter in the simulations and therefore \textsc{turtls} can be used to explore a multitude of different explosion scenarios. During each \textsc{turtls} simulation, \textsc{tardis} \citep{tardis} is used to calculate opacities throughout the model. One of the limitations of \textsc{turtls} is the assumption of local thermodynamic equilibrium (LTE) and therefore simulations are stopped 30~days after explosion. We note however that, by comparing spectra calculated using LTE and non-LTE ionisation approximations, \cite{shen--21} have recently demonstrated that the differences between the two ionisation treatments are negligible until approximately two weeks after maximum light. Therefore the assumption of LTE should have a minimal impact on the light curves of our models. Nevertheless, the model light curves used in this work do not cover the full decline after maximum light. While this study primarily focuses on the recovery efficiency during the earliest phases, and not around maximum light, we also assess the impact of this limitation. We calculate additional survey simulations using SN~Ia templates from \cite{hsiao--07}, which represents the mean evolution of the SN~Ia spectral energy distribution (SED), based on the compilation of a large sample of observed spectra. This is discussed further in Sect~\ref{sect:efficiency}.

\par

For the first scenario considered here, \nifs{} mixing in Chandrasekhar-mass explosions, we use the models provided by \cite{magee--18, magee--20}. These models have been previously shown to broadly reproduce the light curve shapes and colours of multiple SNe~Ia without flux excesses at early times \citep{magee--20, deckers--22}. They may therefore be considered as representative examples of relatively `normal' SNe~Ia and will serve as useful reference points in comparison to models that do produce flux excesses. Each model is characterised by the \nifs{} mass, \nifs{} distribution, and density profile. The shape of the density profile is determined by the kinetic energy, while the \nifs{} distribution is given using the following functional form for the \nifs{} mass fraction at mass coordinate $m$:
\begin{equation}
\label{eqn:ni_dist}
^{56}{\rm Ni}\left(m\right) = \frac{1}{\exp\left(s\left[m - M_{\rm{Ni}}\right]/M_{\rm{\odot}}\right) + 1 },
\end{equation}
where $M_{\rm{Ni}}$ is the total $^{56}$Ni mass in $M_{\odot}$ and $s$ is the scaling parameter, controlling how quickly the ejecta transitions from \nifs{}-rich to \nifs{}-poor. Both the density profile and \nifs{} distribution have a significant impact on the early light curve. \cite{magee--20} showed how the \nifs{} distribution of SNe~Ia can vary greatly, but the vast majority of SNe~Ia considered in their sample were reasonably well matched by intermediate to extended \nifs{} distributions. None of the SNe~Ia considered were reproduced by models in which there is a sharp transition between \nifs{}-rich and \nifs{}-poor regions of the ejecta. In the parameterisation used by \cite{magee--20}, these models are given by large $s$ values (e.g. $s$ = 21, 100), while smaller $s$ values indicate a more extended distribution (e.g. $s = 3, 4.4$), with \nifs{} present throughout the entire ejecta. Given the disagreement between the highly stratified models and observations of SNe~Ia, we do not consider such models as part of our simulations here. For the present work, we take the models with 0.5~\mass{} of \nifs{} used by \cite{deckers--22}, which are consistent with the peak luminosities of SNe~Ia from ZTF \citep{deckers--22} and other studies (e.g. \citealt{scalzo--14}). In this case, for $s = 3$, the \nifs{} fraction in the outermost ejecta is $\sim$0.06, while for $s = 9.7$ it is approximately two orders of magnitude lower (see also fig.~1 of \citealt{magee--20}).

\par

In addition to \nifs{} mixing, we also consider the impact of interaction with a companion red-giant or main sequence star for Chandrasekhar-mass explosions. For this purpose, we take a subset of the previously described \nifs{} mixing models (see Table~\ref{tab:models}) as the base models. For each of these models, we add the flux produced by companion interaction. We use the analytical expressions presented by \cite{kasen--10} (equations 22 and 25) to calculate the black-body emission for a given binary separation ($a$). We note that a few simplifying assumptions are made for these analytical expressions, particularly with regards to the shape of the density profile and the opacity. Nevertheless, they are still useful as an indication of the luminosity expected, are generally consistent with more detailed numerical simulations \citep{kasen--10}, and have been used in numerous studies to search for signs of interaction (e.g. \citealt{hayden--10, brown--12, cao--15, olling--15, miller--19yvq}). For each underlying \nifs{} mixing model, following from \cite{kasen--10}, we consider four types of companions: a $\sim$1~\mass{} red giant ($a$ = 2$\times10^{13}$~cm), a $\sim$6~\mass{} main sequence star ($a$ = 2$\times10^{12}$~cm), a $\sim$2~\mass{} main sequence star ($a$ = 5$\times10^{11}$~cm), and a $\sim$1~\mass{} main sequence star ($a$ = 3$\times10^{11}$~cm). In addition, \cite{brown--12} present an analytical expression for the variation of the black-body emission as a function of viewing angle (see their equation 3). For a given underlying \nifs{} mixing model and companion, we calculate the emission for four viewing angles: 0\degree, 45\degree, 90\degree, and 135\degree. Models observed at 0\degree\, are viewed directly along the line of sight of the shocked region and therefore will show the strongest signs of interaction. Those viewed at 180\degree\, however, would not show signs of interaction and therefore are the same as models without an additional black-body component.

\par

Despite arising from a different mechanism, the early excesses produced by double detonation explosions may be qualitatively similar to those of companion interaction \citep{maeda--18}. In double detonation models, the observables are strongly affected by the post-explosion composition of the helium shell (e.g. \citealt{kromer--10}). For those models containing large amounts of iron-group elements and short-lived radioactive isotopes (in the form of \nifs{}, $^{52}$Fe, or $^{48}$Cr) an excess of flux is produced at early times \citep{jiang--2017, noebauer-17}, while observables around maximum light are typically characterised by red colours and spectra that are inconsistent with normal SNe~Ia \citep{hoeflich--96a, hoeflich--96b, kromer--10, woosley--11}.  \cite{magee--21} present a series of sub-Chandrasekhar mass double detonation models exploring the impact of the helium shell mass and composition on the synthetic observables. For this work, we take those models from \cite{magee--21} with \nifs{}, $^{52}$Fe, or $^{48}$Cr dominated shells. In all cases, a flux excess is produced at early times, the strength and duration of which depends on both the mass and composition of the helium shell. Conversely, models that do not contain iron-group elements in the shell, but are instead dominated by intermediate-mass elements (such as $^{32}$S), do not produce an excess. Instead, these models are generally consistent with normal SNe~Ia \citep{kromer--10, townsley--19, magee--21}. We therefore also include models from \cite{magee--21} with $^{32}$S-dominated shells, as further examples of normal SNe~Ia without flux excesses.

\par

Our final scenario, \nifs{} clumps in the outer ejecta following Chandrasekhar-mass explosions, is qualitatively similar to double detonations in which the post-explosion helium shell is dominated by \nifs{}. A clump of \nifs{} in the outer ejecta was suggested as one possible explanation for the flux excess observed in SN~2018oh \citep{dimitriadis--19}. Following from this, \cite{magee--20b} investigate \nifs{} clumps and their impact on light curves and spectra up to maximum light. \cite{magee--20b} take fiducial models from \cite{magee--20} that reproduce the light curves of SNe~2017cbv and 2018oh around maximum light, but ignore the early flux excesses. To match the shape of the flux excesses, \cite{magee--20b} add a clump of \nifs{} into the outer ejecta of each fiducial model and calculate light curves and spectra. For each fiducial model, they test the range of clump masses and distributions required to match the early observations (see fig.~2 of \citealt{magee--20b}). For our survey simulations, we include all models with \nifs{} clumps from \cite{magee--20b} and extend this to include additional smaller mass \nifs{} clumps calculated as part of this work. Furthermore, \cite{magee--20b} compare their clump models to those with different \nifs{} distributions. We also include these models, and the fiducial SNe~2017cbv and 2018oh models without excesses, in our survey simulations. Again, these models represent normal SNe~Ia that do not produce flux excesses at early times.

\par

\begin{figure}
\centering
\includegraphics[width=\columnwidth]{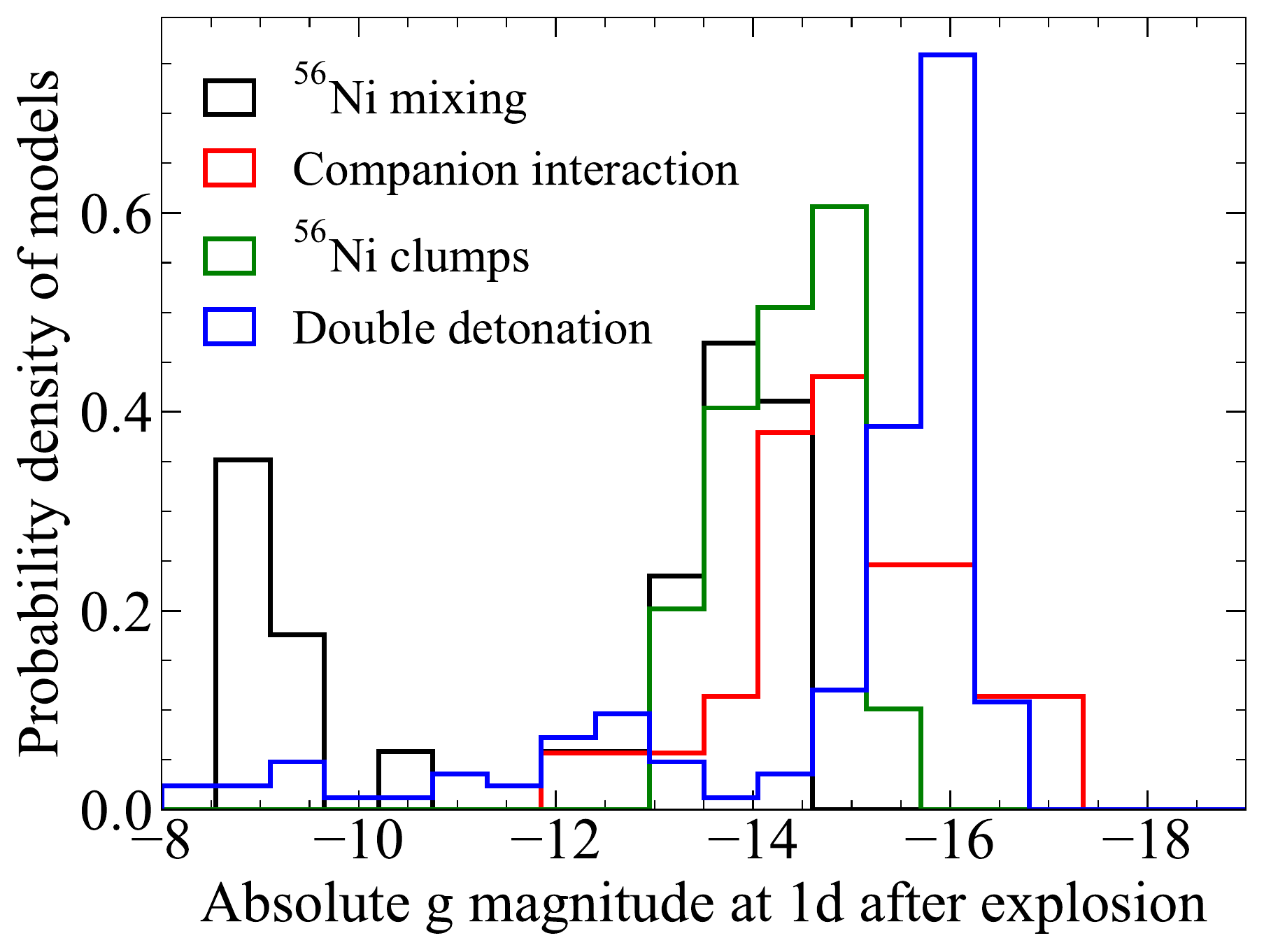}
\caption{Distributions of absolute $g$-band magnitudes for input models at 1\,d after explosion.
}
\label{fig:peaks}
\centering
\end{figure}

Across each of these four scenarios, a total of 313 models are used in this work. This includes 31 \nifs{} mixing models, 96 companion interaction models, 168 double detonation models, and 18 \nifs{} clump models. In Fig.~\ref{fig:peaks}, we show the distribution of absolute $g$-band magnitudes at 1\,d after explosion for each of our models. While these models cover a range of light curve behaviours, the full evolution of each underlying model itself is not overly important. The most relevant aspect for this work is the behaviour of each model within the first few days after explosion. Again, the models included as part of our simulations were chosen due to the broad range of early light curves predicted.

\subsection{Early excess classification}
\label{sect:excess_class}

\begin{figure*}
\centering
\includegraphics[width=\textwidth]{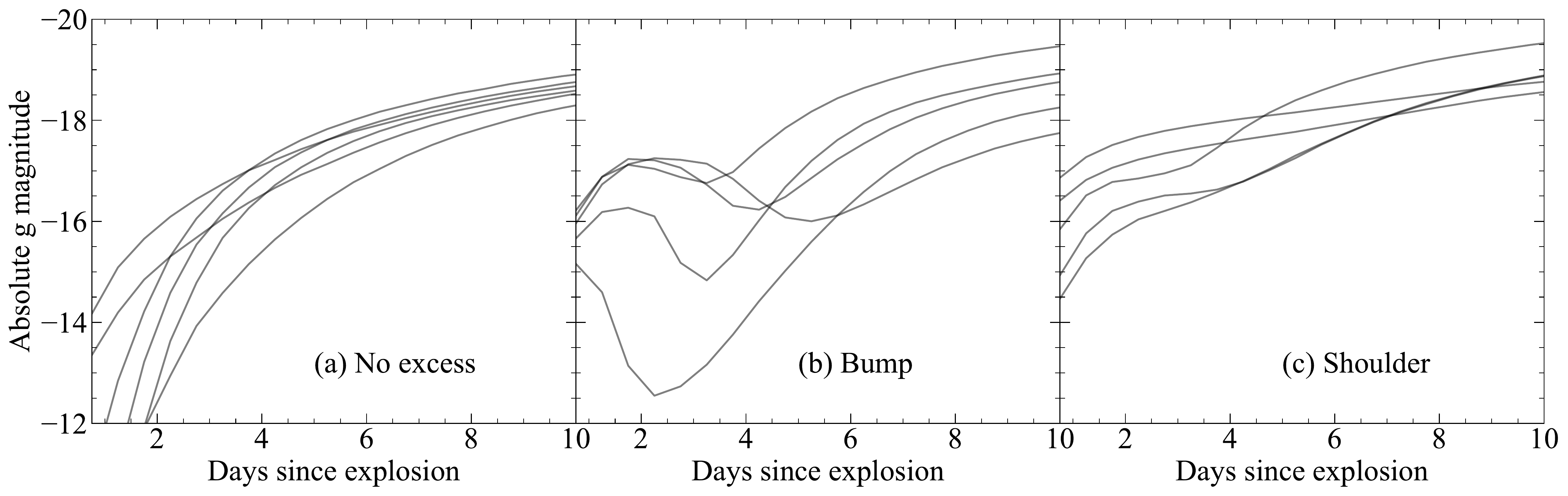}
\caption{Representative examples of early flux excess classifications. Models are classified as having bumps if they show a decline of $\textgreater$0.1~mag after an initial peak. Otherwise models with early excesses are classified as having shoulders.
}
\label{fig:excess_class}
\centering
\end{figure*}

The models included in our survey simulations show significant variation in their early light curves. To aid in the interpretation of our results, it is useful to classify each model based on this behaviour. We therefore present three classifications: no excess, bump, or shoulder. 

\par

Our first classification, no excess, is shown in Fig.~\ref{fig:excess_class}(a). Of our 313 models, 70 would not be expected to contain an excess. This includes all 31 \nifs{} mixing models (as demonstrated by \citealt{magee--20}, a monotically decreasing \nifs{} distribution towards the outer ejecta will not produce an early excess) and the 39 double detonation models with $^{32}$S-dominated shells. The latter are clearly seen in Fig.~\ref{fig:scenario_examples}(c).

\par

For the remaining 243 models, some show an early peak followed by a sharp decline (Fig.~\ref{fig:excess_class}(b)), while others exhibit a quick rise initially followed by a small plateau before joining the main rise of the light curve (Fig.~\ref{fig:excess_class}(c)). These two examples demonstrate the need for classification, as models with an early decline may typically be fainter at early times than those with a plateau. This could make them generally more difficult to detect early. To separate these cases, we define models as containing `bumps' if they show a decline of at least 0.1~mag within the first five days after explosion. The choice of 0.1~mag is arbitrary, but does not have an effect on our results. This classification scheme is simply made to better understand our results, but does not impact our survey simulations. Models that do contain an excess and do not decline by 0.1~mag are classified as having `shoulders'. We also note that this classification scheme is based on models (i.e. \textit{a priori} knowledge of the underlying light curve) and cannot necessarily be easily transferred to observations. In particular, discriminating between a broad light curve, a bump, and a shoulder in real SNe~Ia may be difficult due to limited data and incomplete sampling (e.g. a typical 3\,d cadence). Representative examples of our three early light curve classifications are given in Fig.~\ref{fig:excess_class}.

\begin{table}
\centering
\caption{Model properties}\tabularnewline
\label{tab:models}\tabularnewline
\begin{tabular}{cc}\hline
\hline
Free parameter  & Values studied      \tabularnewline
\hline
\multicolumn{2}{c}{\nifs{} mixing$^1$} \tabularnewline
\hline
Kinetic energy (10$^{51}$~erg) & 0.50, 0.60, 0.65, 0.78, \tabularnewline   & 1.10, 1.40, 1.68, 1.81,\tabularnewline
& 2.18$\phn\phn\phn\phn\phn\phn\phn\phn\phn\phn\phn\phn\phn\phn$ \tabularnewline
\nifs{} mass (\mass{})    & 0.5 \tabularnewline
\nifs{} distribution, $s$ & 3, 4.4, 9.7 \tabularnewline
\hline
\multicolumn{2}{c}{Companion interaction$^{1, 2, 3}$} \tabularnewline
\hline
Kinetic energy (10$^{51}$~erg) & 0.65, 1.10, 1.68\tabularnewline
\nifs{} mass (\mass{})    & 0.5 \tabularnewline
\nifs{} distribution, $s$ & 3, 9.7 \tabularnewline
Companion separation, $a$ (cm) & $2\times10^{13}$, $2\times10^{12}$, \tabularnewline
 & $5\times10^{11}$, $3\times10^{13}$\, \tabularnewline
Viewing angle             & 0\degree, 45\degree, 90\degree, 135\degree \tabularnewline
\hline
\multicolumn{2}{c}{Double detonation$^4$} \tabularnewline
\hline
Core mass (\mass{})  & 0.9, 1.0, 1.1, 1.2 \tabularnewline
Helium shell mass (\mass{})  & 0.01, 0.04, 0.07, 0.10\tabularnewline
Fraction of shell burned  & 0.2, 0.5, 0.8 \tabularnewline
Helium shell composition  & $^{32}$S, $^{48}$Cr, $^{52}$Fe, $^{56}$Ni \tabularnewline
\hline
\multicolumn{2}{c}{\nifs{} clump$^5$} \tabularnewline
\hline
Core \nifs{} mass (\mass{}) & 0.6, 0.8\tabularnewline
Kinetic energy (10$^{51}$~erg) & 1.68, 1.10\tabularnewline
Core \nifs{} distribution, $s$ & 4.4, 9.7 \tabularnewline
\nifs{} clump mass (\mass{}) & 0.000, 0.005, 0.010,\tabularnewline
 & 0.020, 0.030, 0.040\,\tabularnewline
\nifs{} clump width (\mass{}) & 0.06, 0.18\tabularnewline
\hline
\hline
\multicolumn{2}{l}{Further details for each model scenario are provided in the following } \tabularnewline
\multicolumn{2}{l}{references: $^1$ \cite{magee--20}; $^{2}$ \cite{kasen--10}; $^{3}$ \cite{brown--12}; } \tabularnewline
\multicolumn{2}{l}{$^4$ \cite{magee--21}; $^{5}$ \cite{magee--20b} }  \tabularnewline
\end{tabular}
\end{table}

%

\section{simsurvey}
\label{sect:simsurvey}

We use \texttt{simsurvey} \citep{simsurvey} throughout our analysis. \texttt{simsurvey} is a publicly available python package for simulating arbitrary survey schedules and/or transients. It has also been successfully applied to investigating the rates of SNe and kilonovae \citep{simsurvey, de--20, kasliwal--20, almualla--21, andreoni--21, sagues--21}. For each simulation, a survey plan and transient model(s) are required.

\par

The survey plan is given as a list of pointings for each observation and includes the time of observation, right ascension and declination, filter, and limiting magnitude. To demonstrate the impact of survey strategy on the early excess recovery efficiency, we explore a series of toy survey plans for which we arbitrarily select a fixed cadence of 0.2\,d to 7.0\,d and a uniform 5$\sigma$ limiting magnitude of 18 -- 25 in both the $g$- and $r$-bands. The results of our toy simulations are given in Sect.~\ref{sect:toy_sims}.

\par

In addition, we simulate four realistic survey plans based on observing logs from ZTF: ZTF~I, ZTF~I public, ZTF~II public, and high cadence. By using real observing logs from ZTF, these simulations naturally include time lost due to weather and variations in image quality and our predictions can be directly compared with observations. The results of these simulations are presented in Sect.~\ref{sect:ztf_sims}. Observing time for ZTF is split across multiple different public and partnership surveys \citep{ztf--schedules}. Our `ZTF~I' survey plan is based on approximately the first year of ZTF operations (March 2018 -- December 2018). During this time, almost 40\% of the total observing time was allocated to a public 3\,d cadence survey of the northern sky. A large fraction of the partnership observing time was allocated to an extragalactic high cadence survey, with six visits every night for selected fields. Further details of the surveys constituting ZTF observing time are given in \cite{ztf--schedules}. In ZTF~I we include all of the public and partnership observations (but exclude proprietary Caltech time). The `ZTF~I public' survey plan is formed from a subset of the ZTF~I plan and includes only observations from the public survey. Since the end of 2020, the cadence of the ZTF public survey has increased to 2\,d. Therefore `ZTF~II public' includes the first six months of observations from this higher cadence public survey. Finally, the `high cadence' survey plan is also formed from a subset of the ZTF~I survey plan. Here, we select the most well-observed fields and include all observations from the public and partnership surveys. For these fields, the typical cadence is significantly higher at 0.5\,d. Details of each survey strategy are given in Table~\ref{tab:survey_plans}, including the area, length of time covered by our simulations, and median cadence per field.

\par

In addition to the survey plan, we also require a transient generator, which produces the observed light curves based on an underlying model SED, extinction, and redshift. As the relative rates of SNe~Ia with early excesses are unknown, in each survey simulation we assume that only one type of transient will occur (using the models described in Sect.~\ref{sect:model_lcs}). For our toy simulations we focus on a specific subset of the models, which is discussed in Sect.~\ref{sect:toy_sims}. For our more realistic ZTF simulations, we simulate all of the 313 models for each survey plan. Transient coordinates are randomly selected from the area of each survey plan. We include Milky Way extinction, based on the galactic extinction map from \cite{schlegel--98}, and host extinction following from \cite{simsurvey}, where $R_{\rm{V}} = 2$ and $E(B-V)$ is randomly selected from an exponential distribution with a rate of $\lambda$ = 0.11. The time of explosion for each transient is randomly sampled from the duration of each survey plan. Finally, transients are injected up to redshift $z = 0.1$, beyond which the recovery efficiency decreases significantly. In each simulation, we generate 100\,000 transients and calculate the fraction recovered. This results in approximately 31 million transients simulated per ZTF survey plan. 

\par

Throughout this work, we use `detections' to refer to simulated observations in which the flux is greater than the flux uncertainty and inconsistent with zero, for example 5$\sigma$ detections. We use `recovered' to refer to any transient that has at least two 5$\sigma$ detections at any point in the light curve across all bands or meets the stricter criteria described in later sections (e.g. Sect.~\ref{sect:rates}). 

\begin{table}
\centering
\caption{Survey plans used for ZTF simulations}\tabularnewline
\label{tab:survey_plans}\tabularnewline
\resizebox{\columnwidth}{!}{
\begin{tabular}{lcccc}\hline
\hline
Survey plan & Area & Duration & Cadence (days) &   \tabularnewline
\hline
ZTF I        			& 24\,000~deg$^2$ 		& Mar. 2018 -- Dec. 2018	 & 3.0   \tabularnewline
ZTF I public        	& 24\,000~deg$^2$ 		& Mar. 2018 -- Dec. 2018	 & 3.0   \tabularnewline
ZTF II public   		& 24\,000~deg$^2$ 		& Dec. 2020 -- May 2021	 & 2.0    \tabularnewline
High cadence  		& $\phn\phn$110~deg$^2$ 	& Apr. 2018 -- July 2018 & 0.5   \tabularnewline
\hline
\hline
\end{tabular}
}
\end{table}

%

\section{Toy simulations}
\label{sect:toy_sims}

\begin{figure}
\centering
\includegraphics[width=\columnwidth]{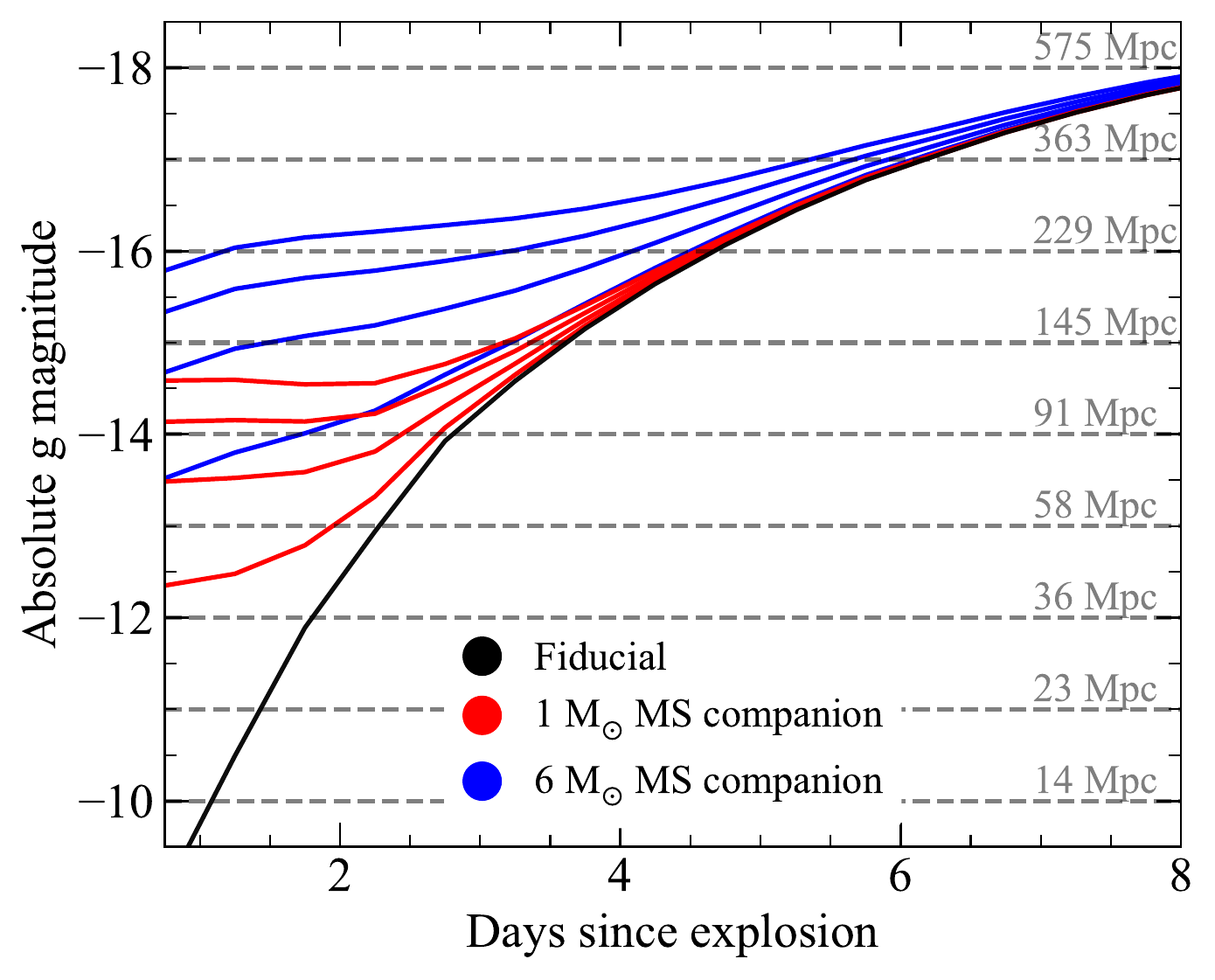}
\caption{Input model light curves for our fiducial model and models to which companion interaction has been added. For each companion, we show four different viewing angles, as given in Table~\ref{tab:models}. As the viewing angle increases, the companion interaction signature decreases. Dashed horizontal lines and values in grey give the maximum luminosity distance at which each absolute magnitude is visible, assuming a 5$\sigma$ limiting magnitude of $m_{\rm{g}} = 20.8$.}
\label{fig:direct_comp}
\centering
\end{figure}

We first demonstrate the importance of survey depth and cadence in the recovery efficiency of young SNe~Ia using a series of toy simulations. We also show how changes in the early luminosity (for example the presence of an excess) affect these recovery efficiencies. For these toy simulations, we assume uniform cadence and survey depth for the $g$- and $r$-bands. Although unrealistic, this allows us to directly demonstrate the improvements that may be expected for higher cadence and/or fainter observations.

\par

As input for our toy simulations, we take a fiducial model without an early excess and compare the early recovery efficiency to models with an excess. As our fiducial model without an excess we select one of our mixing models with an intermediate \nifs{} distribution. For our excess comparisons, we select the models to which the companion interaction signature has been added to the fiducial model. For these purposes, we investigate two types of companions, a 1~\mass{} main sequence star and a 6~\mass{} main sequence star. For both types of companions, we also include the full range of viewing angles given in Table~\ref{tab:models}. The input model light curves are shown in Fig.~\ref{fig:direct_comp}. In Fig.~\ref{fig:direct_comp}, we also show the maximum redshift at which various absolute magnitudes would be visible, assuming a 5$\sigma$ limiting magnitude of $m_{\rm{g}} = 20.8$. From Fig.~\ref{fig:direct_comp}, it is clear that our fiducial model could only be observed at $\sim$1\,d after explosion if it was extremely nearby ($\lesssim$15~Mpc). The signatures of companion interaction with a 1~\mass{} MS companion star increase the early luminosity by 2 -- 4~mag, depending on the viewing angle. As such, the models should be detectable at 1\,d after explosion out to 40 -- 120~Mpc. Likewise, interaction with a 6~\mass{} MS companion increases the early luminosity by 3 -- 5~mag and hence should be detectable out to 80 -- 230~Mpc. Early excesses that increase the luminosity by a factor of $X$ should similarly increase the observable distance by a factor of $\sqrt{X}$.

\par

\begin{figure*}
    \centering
    \begin{subfigure}[b]{\textwidth}
        \includegraphics[width=\textwidth]{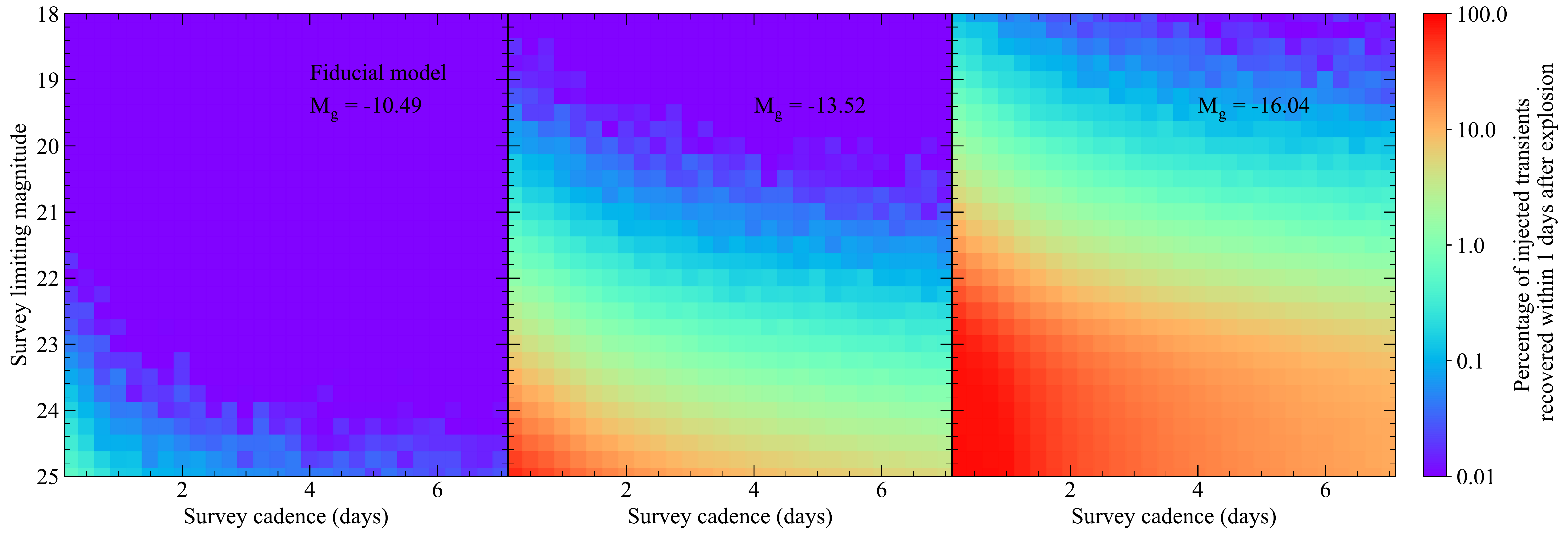}
    \end{subfigure}
    \\
    \begin{subfigure}[b]{\textwidth}
        \includegraphics[width=\textwidth]{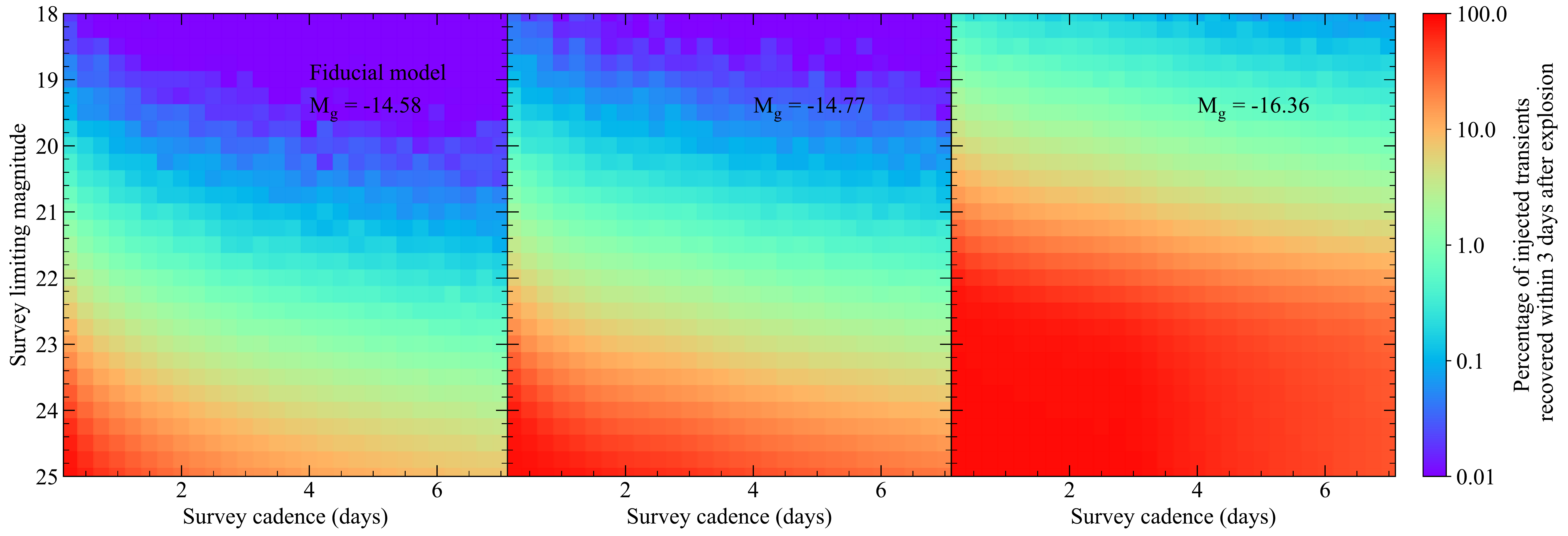}
    \end{subfigure}

    \caption{ Percentage of injected transients recovered within one (top) and three (bottom) days of explosion for arbitrary survey depths and cadences. For each model, we give the absolute $g$-band magnitude at both times. We note that the colour-bar is shown on a log-scale.
   }
    \label{fig:cadence_depth}
\end{figure*}

Figure~\ref{fig:cadence_depth} shows the results of our toy simulations and gives the percentage of injected models that are recovered within 1 and 3\,d of explosion for various combinations of survey depth and cadence. Up to 1\,d after explosion, our fiducial model is not recovered for almost all toy surveys. As shown by Fig.~\ref{fig:direct_comp}, this results from the faint luminosity at this time, which will only be observable for the most nearby cases. With a survey depth of $m \sim 25$, the first $\sim$1\,d of our fiducial model would be observable out to $\sim$100~Mpc, however a high cadence is also required to actually observe these fleeting moments. Even in these cases $\lesssim$1\% of our injected fiducial models are recovered within 1\,d of explosion. Increasing the early luminosity by $\sim$3~mag results in $\sim$0.05\% of SNe~Ia being detected within 1\,d of explosion for a typical depth and cadence of 21 and 3\,d. Further increasing the early luminosity by an additional $\sim$2.5~mag results in $\sim$1.2\% of the injected models being recovered within 1\,d of explosion. As shown by Fig.~\ref{fig:direct_comp}, a typical survey depth of $\sim$21 would be sensitive to these SNe up to a redshift of $\sim$220~Mpc. By 3\,d after explosion, our fiducial model would still only be visible out to $\sim$100~Mpc (see Fig.~\ref{fig:direct_comp}). At this time, $\sim$0.15\% of our injected fiducial models have been recovered for a typical depth and cadence of 21 and 3\,d. In comparison, 0.5 -- 8\% of the excess models have been recovered. Despite being a similar magnitude to the fiducial model at this time, the excess model shown in Fig.~\ref{fig:cadence_depth} was significantly brighter at earlier times and hence more of the injected models have been recovered.

\par

From Fig.~\ref{fig:cadence_depth} the impact of cadence is less significant than survey depth on determining the number of young SNe~Ia discovered. Moving from a three to two day cadence for the same survey depth likewise results in $\sim$50\% more SNe~Ia discovered within approximately 1\,d of explosion. We note however that a reasonably high cadence would be necessary to, for example, resolve the shape of the early excess and provide tighter constraints on pre-explosion non-detections. In Sect.~\ref{sect:ztf_sims}, we compare the results of our toy simulations to the more detailed simulations of ZTF. In addition, these toy simulations assume uniform cadence for both bands and therefore may not necessarily be applicable to surveys with more complicated scheduling. For example, the upcoming Vera C. Rubin Observatory's Legacy Survey of Space \& Time (LSST; \citealt{ivezic--08, lsst--09}) is expected to reach a single-visit depth of $\sim$24 in the $g$- and $r$-bands. Based on Fig.~\ref{fig:cadence_depth}, one may naively expect that $\gtrsim$10\% of all SNe~Ia up to $z = 0.1$ would be discovered by LSST early enough to determine whether an excess is present. The baseline \texttt{minion\_1016} observing strategy has a typical inter-night gap of $\sim$3\,d across all bands, while the gap for same-band observations could be $\gtrsim$15\,d. Since excesses are predicted to be prominent in blue bands and not redder bands, this could result in cases where the SN excess is simply below the detection threshold in redder bands even though the field was observed. Hence, the recovery fractions given in Fig.~\ref{fig:cadence_depth} would be overly optimistic. The proposed \texttt{altsched} observing strategy does have a typical inter-gap for $r$-band observations of $\sim$3\,d, however the gap between $g$-band observations remains large. We therefore caution against over interpretation of Fig.~\ref{fig:cadence_depth} for surveys with complicated scheduling requirements. More detailed simulations are necessary to fully understand the effect of these on the recovery of young SNe~Ia.

%
\section{ZTF simulations}
\label{sect:ztf_sims}
\subsection{General recovery efficiency}
\label{sect:efficiency}

\begin{figure}
\centering
\includegraphics[width=\columnwidth]{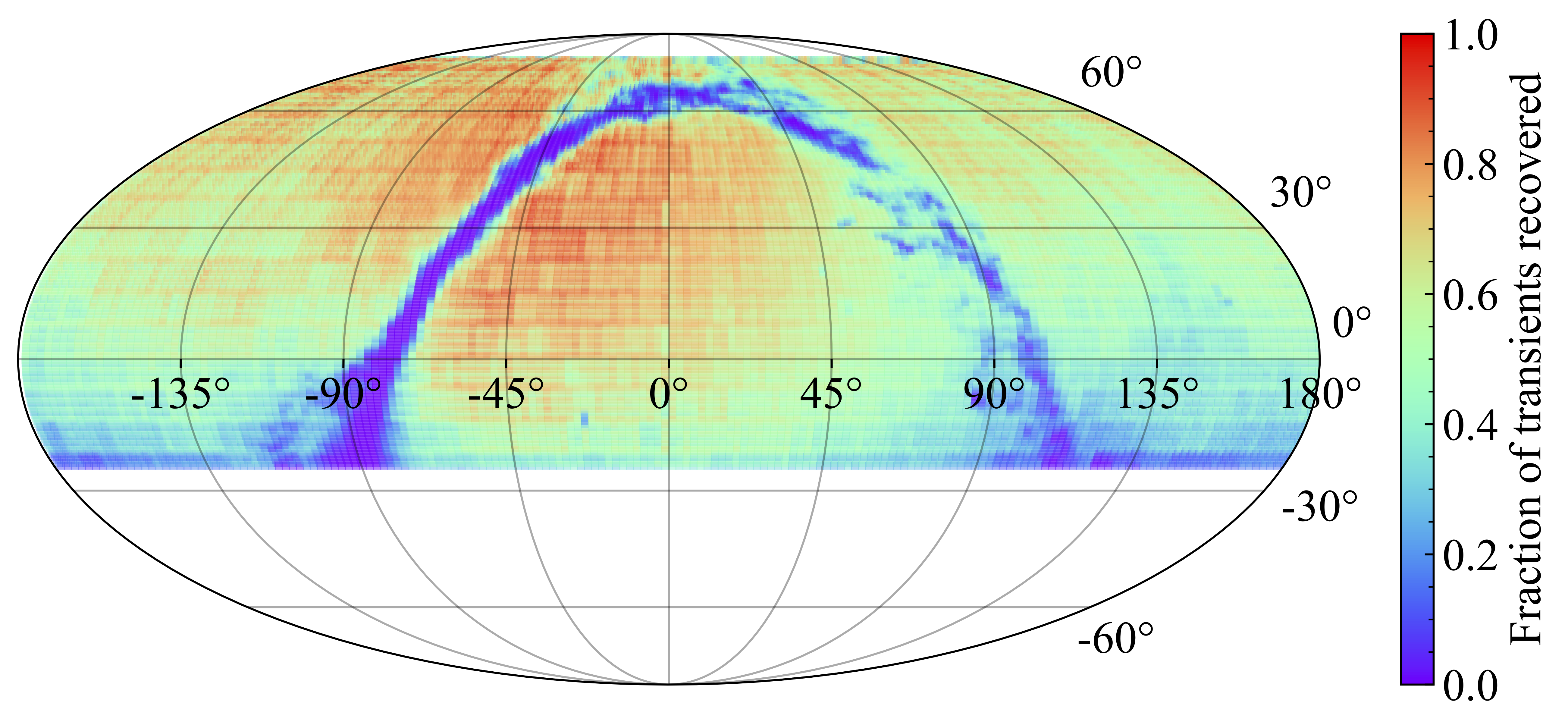}
\caption{Recovery fraction for our ZTF~I survey plan at different positions on the sky.  }
\label{fig:skymaps}
\centering
\end{figure}

Here we briefly discuss the overall recovery efficiency of ZTF.
In Fig.~\ref{fig:skymaps}, we show a sky-map of the recovery fraction for our ZTF~I survey plan, which includes all of the 313 models discussed in Sect.~\ref{sect:model_lcs}. As demonstrated by Fig.~\ref{fig:skymaps}, our ZTF~I (and indeed ZTF~I and II public) survey plan include fields around the galactic plane, which have significantly smaller recovery fractions than other sky positions. We exclude those transients injected close to the galactic plane ($|b|$~\textless~15\degree) from our recovery statistics. 

\par

We find the mean recovery fraction for our ZTF~I survey plan is $\sim$56\%. This accounts for transients that were lost due to occurring in fields not observed when the SN was active and those falling into field/chip gaps, in addition to those that simply were not bright enough to pass the detection criteria. For the ZTF~I and II public survey plans, we find similar recovery efficiencies. Unsurprisingly, the high cadence survey plan has the highest recovery efficiency ($\sim$90\%) as it is focused on a much smaller area and time. Figure~\ref{fig:direct_comp} demonstrates that ZTF should be sensitive to SNe~Ia across the entire redshift range of our simulations ($z \leq 0.1$) for a few weeks surrounding maximum light. As such, we find only minor decreases in the recovery efficiencies at higher redshift compared to low redshift (4 -- 8\%).

\par

\par

As previously mentioned, the light curves used in these simulations extend up to approximately a few days after maximum light. The overall recovery fractions discussed thus far for our ZTF simulations will therefore typically be lower than simulations that include the full light curve evolution. To assess the impact of this limitation, we perform additional simulations using each of our survey plans and the \cite{hsiao--07} spectroscopic template. We use two light curves based on this template; the first extends up to 40\,d post-maximum (full), and the second extends up to only 5\,d post-maximum (early). When using only the early \cite{hsiao--07} template light curve, we find recovery efficiencies comparable to our other ZTF simulations. When considering the full light curve however, there is a $\lesssim$13\% increase in the overall fraction of transients recovered across the redshift range simulated for all of our survey plans. The more limited light curve coverage for the models used in this work therefore has only a minor impact on our ability to recover transients, but the overall recovery efficiencies quoted here should be considered as lower limits with an absolute uncertainty of $\sim$13\%. This does not impact the recovery efficiencies for young SNe~Ia, which are the main focus of this work.

%

%

\subsection{Phase of first detection}
\label{sect:early_detections}

\begin{figure*}
\centering
\includegraphics[width=\textwidth]{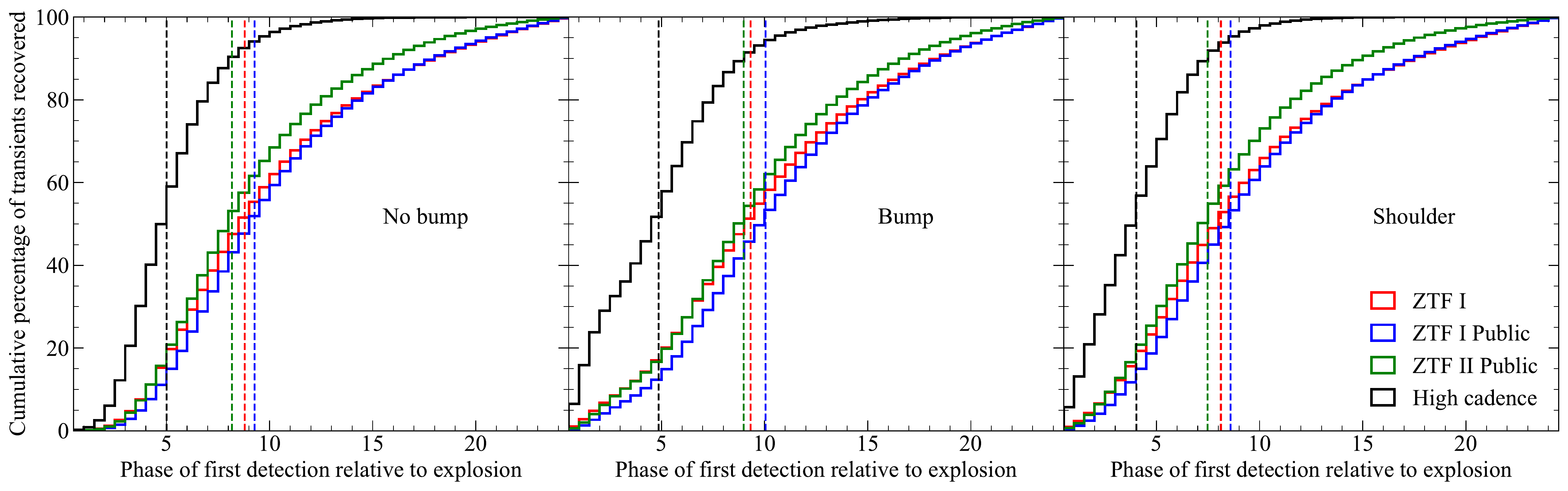}
\caption{Distribution of rest-frame first detection phase for each of our survey plans and early light curve classifications. Vertical dashed lines show the median detection phase for each survey plan. 
}
\label{fig:first_detection}
\centering
\end{figure*}

Figure~\ref{fig:first_detection} shows the cumulative distribution of rest-frame phases at which the injected transients are first detected at 5$\sigma$. This is shown for each of our excess classifications (see Sect~\ref{sect:excess_class}) and survey plans.

\par

From Fig.~\ref{fig:first_detection}, the median detection phase from the high cadence survey plan is 4 -- 5\,d after explosion depending on the early excess classification. Almost all first detections have been made by 7 -- 9\,d after explosion. For the no excess models, $\lesssim$0.2\%(12\%) of detections have been made by 1\,d(3\,d) after explosion. These results are comparable to our expectations based on the toy simulations shown in Fig.~\ref{fig:cadence_depth}. Assuming a typical depth and cadence of 21 and 0.5\,d, for models without excesses 0 -- 1\% of the injected light curves should be recovered within $\sim$1\,d of explosion, depending on their exact light curve shape (see Fig.~\ref{fig:excess_class}). Likewise, 5.5 -- 6.5\%(32 -- 35\%) of first detections occur within 1\,d(3\,d) for the bump and shoulder models. Given the typical luminosities of the bump and shoulder models, our toy simulations indicate that 2 -- 6\% of the light curves are recovered within 1\,d of explosion. This is again comparable to the prediction from our detailed ZTF simulation. Our toy simulations are therefore consistent with the more detailed ZTF simulations to within a factor of approximately a few. Discrepancies likely arise from differences in the specific light curve shapes of the models used in the ZTF simulations compared to the subset shown in Fig.~\ref{fig:cadence_depth}.

\par

The impact of an improved cadence is also demonstrated by the comparison between the first detection distributions of the ZTF I and II public simulations. For ZTF~I public, the median phase of first detection is 8 -- 10\,d after explosion. In the ZTF~II public survey the median detection phase is systematically lower by $\sim$1\,d for each of the light curve types. The change from 3\,d to 2\,d cadence is therefore directly reflected in the phases of recovered transients. Given that the typical lifetimes of excesses are $\lesssim$5\,d however, for both survey plans we would expect the majority of SNe~Ia to show no signs of an excess, even if they all actually had one, as they are simply not observed early enough. Nevertheless, in ZTF I public, $\sim$0.02\%(1.5\%) of first detections for models without excesses occur within 1\,d(3\,d) of explosion, while 0.4 -- 0.5\%(5.6 -- 7.3\%) of first detections have occurred for models with an excess. In ZTF~II public, this increases slightly to $\sim$0.03\%(2.3\%) for no excess and 0.6 -- 0.7\%(8.4 -- 11.1\%) for models with an excess. The improved cadence in ZTF~II is therefore consistent with the prediction from our toy simulations that moving from a 3\,d cadence to 2\,d should result in $\sim$50\% more young SNe~Ia being discovered. Our simulations indicate that the ZTF high cadence survey could show up to approximately an order of magnitude increase in the number of young SNe~Ia discovered relative to a 3\,d cadence survey. While this is slightly higher than expected based on our toy simulations, the high cadence survey plan covered a smaller area and time-span and therefore is less affected by, for example, weather losses than the public surveys. We again note that the model light curves used in this study extend up to only a few days after maximum light. While the absolute number of transients recovered at early times will obviously not change, the relative proportion of these to the total number of recovered transients will. Therefore, given that the overall number of recovered transients is underestimated $\lesssim$13\% (Sect.~\ref{sect:efficiency}), these early recovery fractions should be considered to be upper limits with conservative relative uncertainties of 1 -- 2\%.

\par

Although our simulations have focused on SNe~Ia, these results, and the efficiencies shown in Fig.~\ref{fig:cadence_depth}, should be broadly applicable to other SNe types. Given that the recovery efficiency of SNe~Ia at 1\,d after explosion is approximately 20$\times$ higher for those with an excess compared to those without, we would therefore expect that the majority of SNe~Ia discovered at 1\,d after explosion would contain an excess. This does however assume that SNe~Ia with or without an excess are equally likely and the only limitation is our ability to detect sufficiently early. The fact that samples of young SNe~Ia do not show such an overabundance of excesses indicates that they must instead be relatively rare. This point is discussed further in Sect.~\ref{sect:excess_rate}.

\begin{figure}
\centering
\includegraphics[width=\columnwidth]{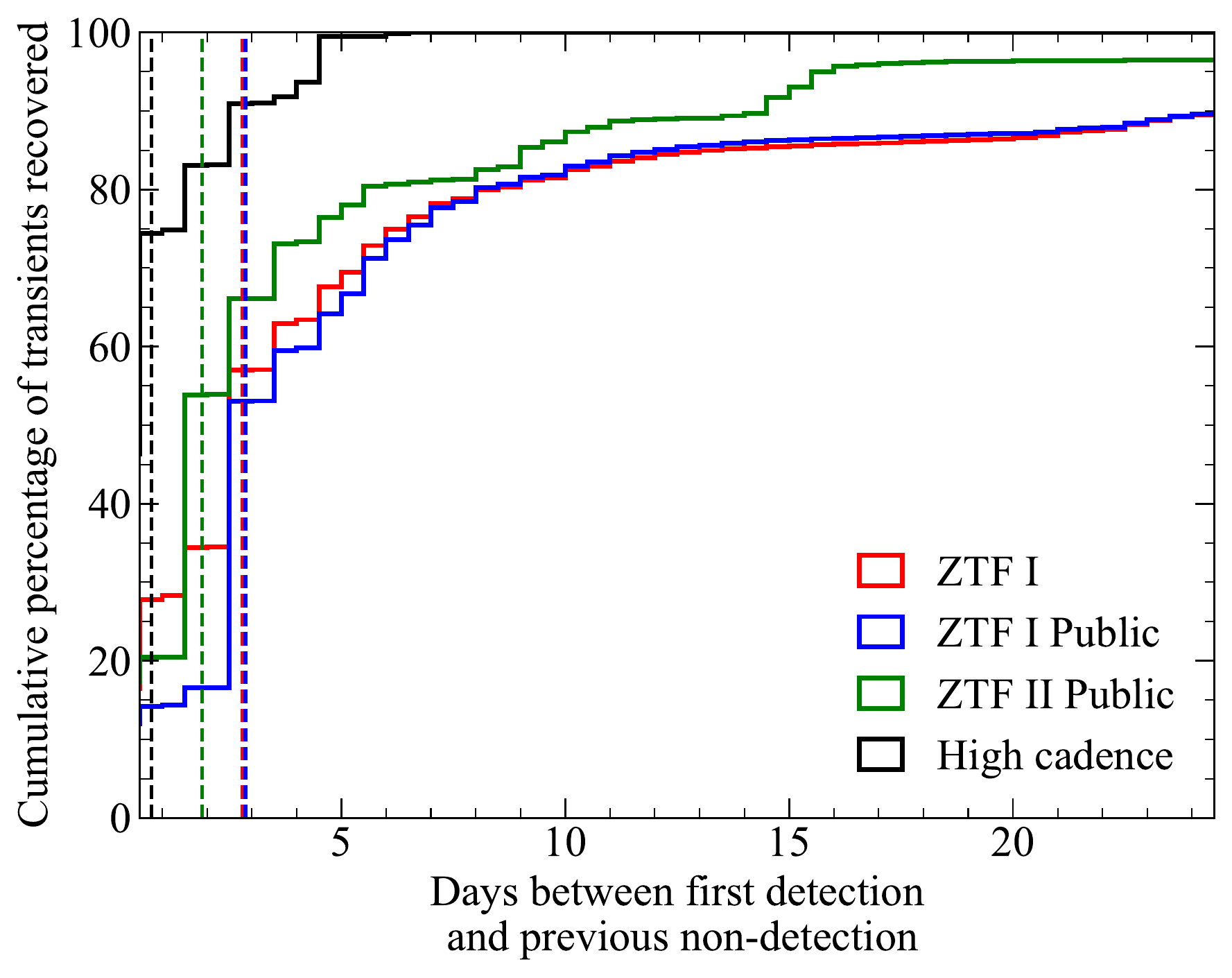}
\caption{Distribution of the times between first detection and previous non-detection. Vertical dashed lines show the median for each survey plan. 
}
\label{fig:non_detection}
\centering
\end{figure}

\par

In Fig.~\ref{fig:non_detection} we show the distribution of non-detection phases for each of our survey plans. Here we define the non-detection phase as the time between the first 5$\sigma$ detection and previous non-detection. For the high cadence survey plan, $\sim$75\% of recovered transients have a non-detection within 1\,d of the first detection, while all transients have previous non-detections within $\sim$5\,d. As it also includes observations of high cadence fields, the ZTF~I survey plan shows a slightly higher fraction of non-detections within 1\,d compared to both public plans, however the median phase of non-detection is $\sim$3\,d. For ZTF~I and II public, the difference in their median non-detection phases (3\,d and 2\,d, respectively) is again representative of the difference in cadence for the survey strategies. For ZTF~I public, $\sim$10\% of all recovered transients do not have a previous non-detection. In ZTF~II public, this drops to $\lesssim$4\%, however this may also be due to the shorter length of the survey simulated.


\subsection{Early detections as a function of absolute magnitude} 
\label{sect:early_mag_v_z}

\begin{figure*}
    \centering
    \begin{subfigure}[b]{\textwidth}
        \includegraphics[width=\textwidth]{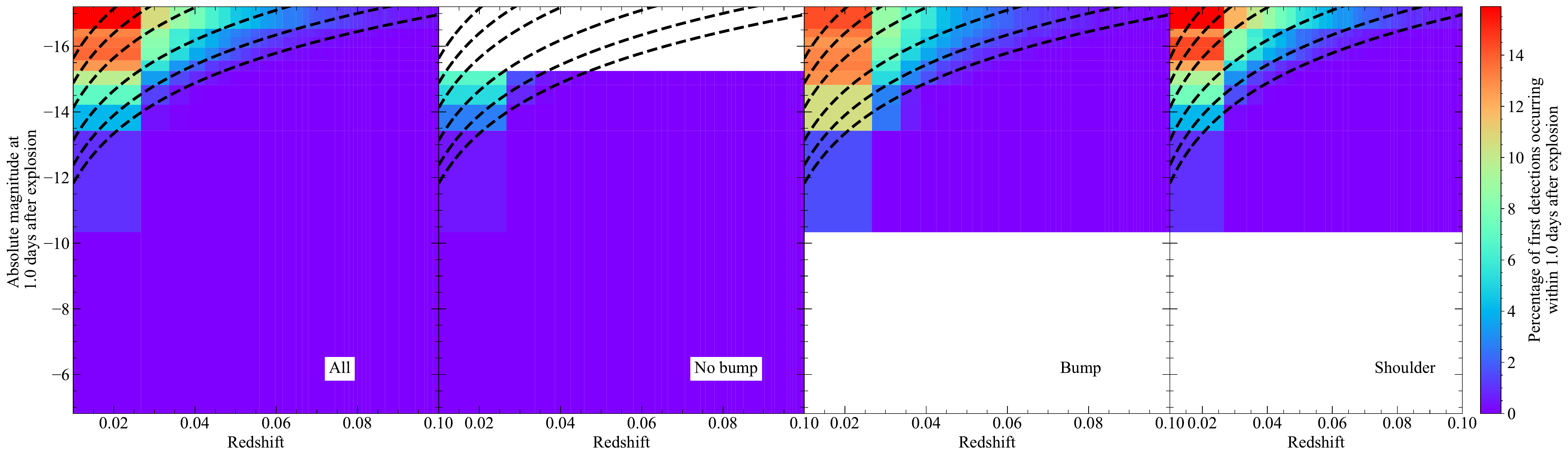}
    \end{subfigure}
    \\
    \begin{subfigure}[b]{\textwidth}
        \includegraphics[width=\textwidth]{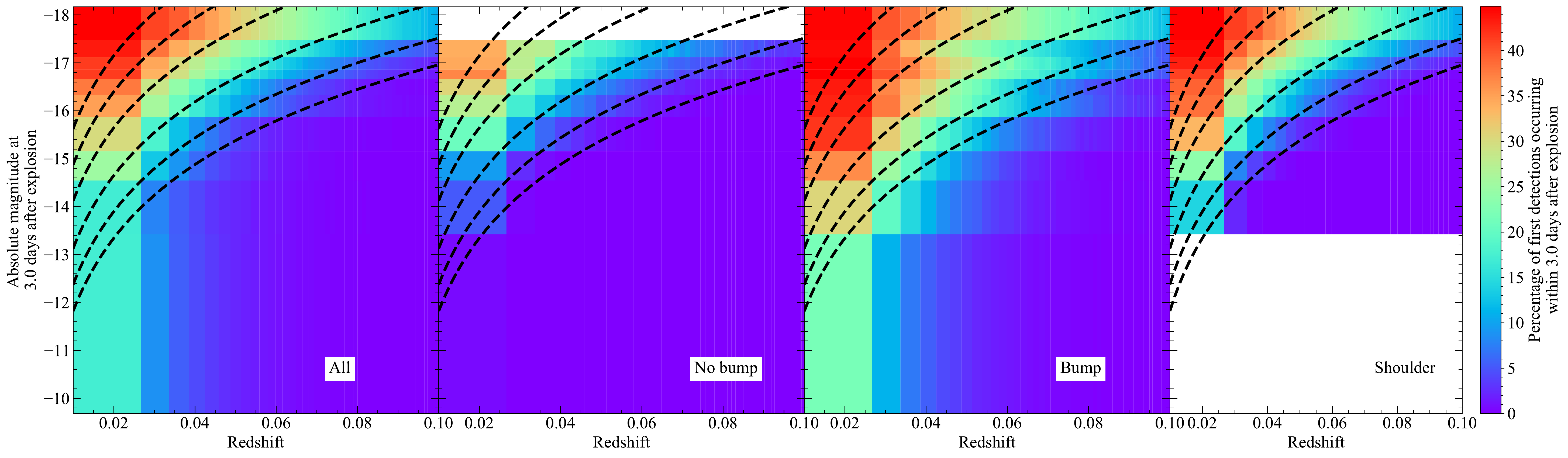}
    \end{subfigure}

    \caption{Fraction of recovered transients from the ZTF~I survey plan detected before certain times as a function of absolute magnitude at that time and redshift. Times shown here are 1\,d and 3\,d post-explosion. In each row, we show the fraction for all transients and each of our early light curve classifications separately. We also note the different scales for colours in each row. Black dashed lines show the redshift at which a given absolute magnitude is observed with signal-to-noise ratios of 3, 5, 10, 25, 50, and 100. We note the different scales for the top and bottom panels.
   }
    \label{fig:y1_mag_v_z}
\end{figure*}

In Fig.~\ref{fig:y1_mag_v_z} we show the impact of both absolute magnitude and redshift on the recovery statistics at different times for our ZTF~I survey plan simulations. Figure~\ref{fig:y1_mag_v_z} directly demonstrates the sensitivity of ZTF to early excesses at various redshifts. We separate all of the recovered transients into discrete, two-dimensional bins across redshift and absolute magnitude. Redshift bins are set to each encompass an equal volume, while the absolute magnitude bins contain an equal number of models. The absolute magnitudes are measured at 1\,d and 3\,d after explosion. With the magnitude and redshift bins defined, we calculate the fraction of recovered transients that were detected before each time within each bin. Figure~\ref{fig:y1_mag_v_z} indicates the redshift at which each absolute magnitude would be observed with various signal-to-noise ratios. Similar plots for our ZTF~I and II public, and high cadence survey plans are shown in Figs.~\ref{fig:y1_pub_mag_v_z}, \ref{fig:ztfii_pub_mag_v_z}, and \ref{fig:hc_mag_v_z} respectively. 

\par

Within 1\,d of explosion, many of the models without an excess are still experiencing the so-called dark phase \citep{piro-nakar-2013}, the time between explosion and the first light emerging from the SN. Hence, none of our models without bumps are brighter than approximately $M_{\rm{g}} = -15.3$. In contrast, none of our models with early excesses are fainter than $M_{\rm{g}} = -10.3$.

\par 

Figure~\ref{fig:y1_mag_v_z} shows that the impact of the different early light curve behaviours on the recovery efficiencies is clearly apparent. The recovery efficiencies follow the expected trends, showing a greater proportion of early detections for SNe brighter at theses phases and at low redshift. For example, within the first redshift bin ($z \lesssim 0.027$), for transients brighter than $M_{\rm{g}} = -15.3$ at 1\,d (i.e. those with an excess) 13 -- 16\% of recovered transients are first detected within 1\,d of explosion. These SNe should be observed with high signal-to-noise ratios of $\gtrsim10$. These percentages steadily decrease out to higher redshifts and drop below $\sim$1\% at $z = 0.05$, where the signal-to-noise ratio is $\gtrsim$3. For somewhat fainter models, those with an excess typically show higher recovery efficiencies than those without. In our first redshift bin, for models with $-15 \lesssim M_{\rm{g}} \lesssim -13.5$, only 3 -- 7\% of detections occur within 1\,d of explosion for models without an excess compared to 8 -- 13\% for models with a bump or shoulder. Although all models are binned here based on their absolute magnitudes at 1\,d after explosion, the increased recovery efficiency of models with an excess is due to their overall faster rise during this time. Hence their average luminosity is higher for times $\lesssim$1\,d after explosion compared to models without an excess, which are likely just exiting the dark phase. We also find that models with early excesses are observable to slightly higher redshifts compared to models without excesses within the same absolute magnitude bins.

\par

For $z \lesssim 0.05$, 32 -- 45\% of all detections occur within 3\,d of explosion for bright models (those that are brighter than $M_{\rm{g}} = -17.5$ at 3\,d after explosion). Even out to the high redshift limit of our simulations, $z = 0.1$, as many as $\sim$11\% of detections occur within 3\,d of explosion for these models. At these redshifts, the signal-to-noise ratio is 5 -- 10. In contrast, for the faintest magnitude bin ($M_{\rm{g}} \gtrsim -13$) and lowest redshift bin, less than 1\% of detections occur within 3\,d of explosion for those models without an excess. The signal-to-noise ratio of these models is $\lesssim$10. This percentage increases to $\sim$22\% for models with bumps.

\par

Comparing Figs.~\ref{fig:y1_pub_mag_v_z} \& \ref{fig:ztfii_pub_mag_v_z}, we again show that a $\sim$50\% increase in the number of young SNe~Ia recovered should be expected for the higher cadence ZTF~II public survey. For low redshift, bright SNe Ia, typically $\sim$10\%(39\%) of detections occur within 1\,d(3\,d) of explosion for models with an excess in the ZTF~I public survey, while within ZTF~II public this increases to $\sim$16\%(55\%). Similarly for the bright models without an excess at low redshift, $\sim$4\%(28\%) of detections should occur within 1\,d(3\,d) of explosion in ZTF~I public, compared to $\sim$6\%(42\%) for ZTF~II public.

\subsection{Summary}

Our simulations demonstrate the relative proportions of young SNe~Ia that should be detected by ZTF and provide quantified recovery efficiencies to assess the likelihood of discovery. Of those transients recovered by our simulations, a significantly higher fraction of detections are made within the first few days of explosion for models with an excess compared to those without. Although the number of young SNe~Ia discovered by surveys is somewhat small, there is no such strong preference for those with excesses. This would indicate that flux excesses (at least those as bright as predicted by simulations) must be relatively rare. The comparison between our ZTF~I and II public survey plans also shows that the improved cadence of ZTF~II should produce a $\sim$50\% increase in the number of young SNe discovered. Similarly, the ZTF high cadence survey should result in approximately up to an order of magnitude increase in the number of young SNe discovered. While the area of the high cadence survey is also approximately an order of magnitude smaller than the public surveys, the significantly improved temporal resolution should allow for the shapes of the early light curves to be better resolved and hence provide more meaningful constraints on the early excess. Although our simulations have focused on models of SNe~Ia, we expect that this is also broadly applicable to other SNe types.

%


%

\section{Rates}
\label{sect:rates}

Given the recovery efficiencies described in Sect.~\ref{sect:ztf_sims}, we can now place limits on the contribution of SNe~Ia with excesses to the overall SNe~Ia rate. We first verify the reliability of our simulations by applying our efficiencies to calculating the overall rate of SNe~Ia. 

\par

\subsection{Rate of SNe~Ia}
\label{sect:ia_rate}

To calculate the rate of normal SNe~Ia, we follow a similar method to that described by \cite{frohmaier--19} for PTF, a precursor to ZTF. The rate of SNe~Ia is defined as
\begin{equation}
    r_v(z)=\frac{1}{V \Delta T}\sum_{i=1}^{N} \frac{1+z_i}{\epsilon_i},
    \label{eqn:rate}
\end{equation}
which counts the number of SNe~Ia, $N$, that exploded within a given volume $V$ and timespan $\Delta T$. Time dilation effects for each SN redshift are accounted for by the inclusion of the factor $1 + z_i$. Each SN is also weighted by the recovery efficiency, $\epsilon$, which accounts for any similar SNe~Ia that simply were not detected by the survey. The volume $V$ is given by
\begin{equation}
V=\frac{\Theta}{41253}\frac{4\pi}{3}\left[\frac{c}{H_0} \int_{z_1}^{z_2} \frac{dz}{\sqrt{\Omega_M(1+z)^3+\Omega_\Lambda )}}\right ]^3\textrm{Mpc}^3,
\label{eqn:volume}
\end{equation}
where $z_1$ and $z_2$ give the redshift boundaries over which the rate will be calculated and $\Theta$ gives the survey area in deg$^2$. Here, we assume $\Omega_M$ = 0.3, $\Omega_{\Lambda}$ = 0.7, and $H_0$ = 70~\kms{}~Mpc$^{-1}$.

\par

We use the sample of ZTF SNe~Ia presented by \cite{yao--19} to calculate the SNe~Ia rate. This sample includes 127 SNe~Ia observed during the partnership high cadence survey that were discovered at very early times (more than ten rest-frame days before $B$-band maximum) and therefore will form the basis for our estimates of the fraction of SNe~Ia with excesses. As our simulations only include redshifts up to $z = 0.1$, we exclude higher redshift SNe from the following rate calculation, leaving a sample size of 90.

\par

For the remaining SNe, we assign a weight based on the product of the recovery efficiency and spectroscopic completeness for the SN redshift. We determine the recovery efficiency by applying the same selection criteria used by \cite{yao--19} (see their table~1) to our high cadence survey plan simulations calculated with the full \cite{hsiao--07} template light curve. These selection criteria are therefore stricter than those used thus far (two 5$\sigma$ detections at any time), but as they are the same as what was applied to the observed sample they allow for a direct comparison between simulations and observations. As part of their criteria \cite{yao--19} also require that reference images for each SN are obtained earlier than 25\,d before $B$-band maximum light. Approximately 81\% of SNe meet this specific criterion. As our simulations do not include reference images, we assume a $\sim$81\% likelihood that each SN would pass this selection cut and adjust the recovery efficiency accordingly. The spectroscopic completeness is given by \cite{fremling--20} and \cite{perley--20}, who estimate that ZTF is $\gtrsim$95\% spectroscopically complete down to $m_{\rm{peak}} \textless$18.5 and 75 -- 89\% complete at $m_{\rm{peak}} \textless$19. We convert this to a redshift dependence, assuming a typical absolute magnitude for SNe~Ia of $M_{\rm{g}} = -19.2$.

\par

Summing over the 90 SNe~Ia in the sample and appropriate weightings, as in Eqn.~\ref{eqn:rate}, we calculate a SN~Ia rate for this sample of $r_v = 2.56\pm0.24$(stat)$\pm0.06$(sys)$\times 10^{-5}$~SNe~yr$^{-1}$~Mpc$^{-3}$. We estimate the statistical uncertainty as proportional to $\sqrt{N}$, where $N$ = 90. Here the systematic uncertainty includes only the uncertainty resulting from spectroscopic completeness variations. For the PTF SNe~Ia sample, \cite{frohmaier--19} calculate a SN~Ia rate of $r_v = 2.43\pm0.29$(stat)$^{+0.33}_{-0.19}$(sys)$\times 10^{-5}$~SNe~yr$^{-1}$~Mpc$^{-3}$. The agreement between both rates estimates demonstrates the reliability of our simulations and how the same selection criteria applied to our simulations results in a consistent SN~Ia rate. We may therefore apply new cuts to both our simulations and the \cite{yao--19} sample that will allow us to directly estimate the fraction of SNe~Ia with an excess.

\begin{figure*}
\centering
\includegraphics[width=\textwidth]{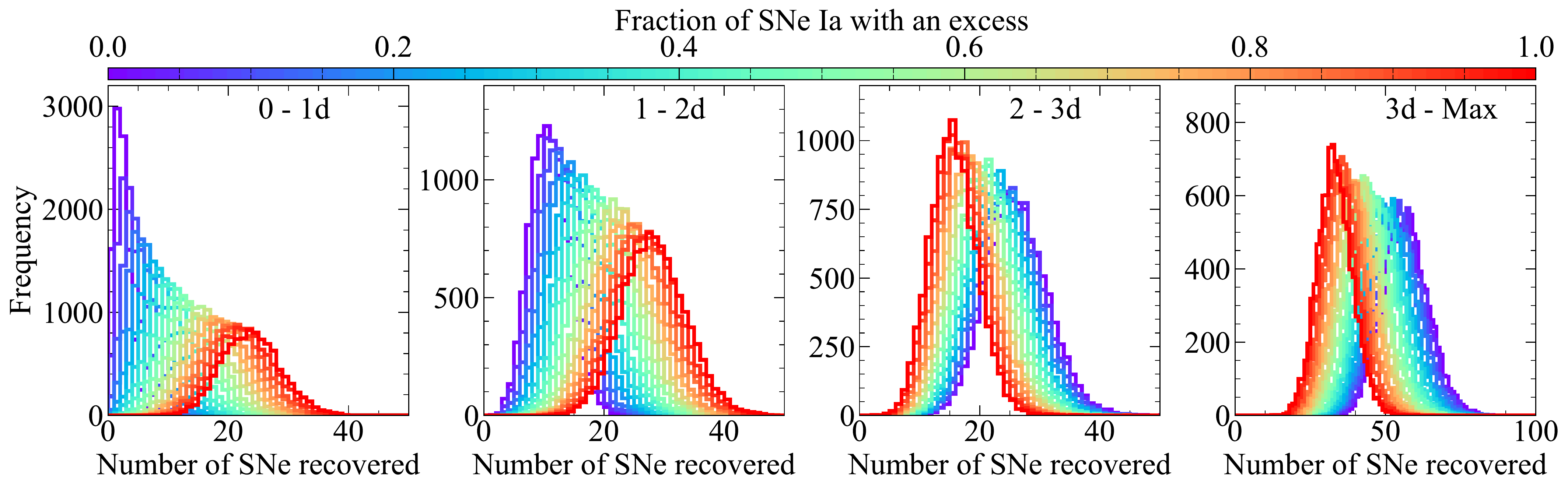}
\caption{ Number of SNe~Ia recovered within various time bins, assuming a given fraction contain an excess at early times. For each fraction, we show the distribution in the number recovered across 10\,000 survey simulations with randomly sampled input models. 
}
\label{fig:counts_dist}
\centering
\end{figure*}

\begin{figure*}
\centering
\includegraphics[width=\textwidth]{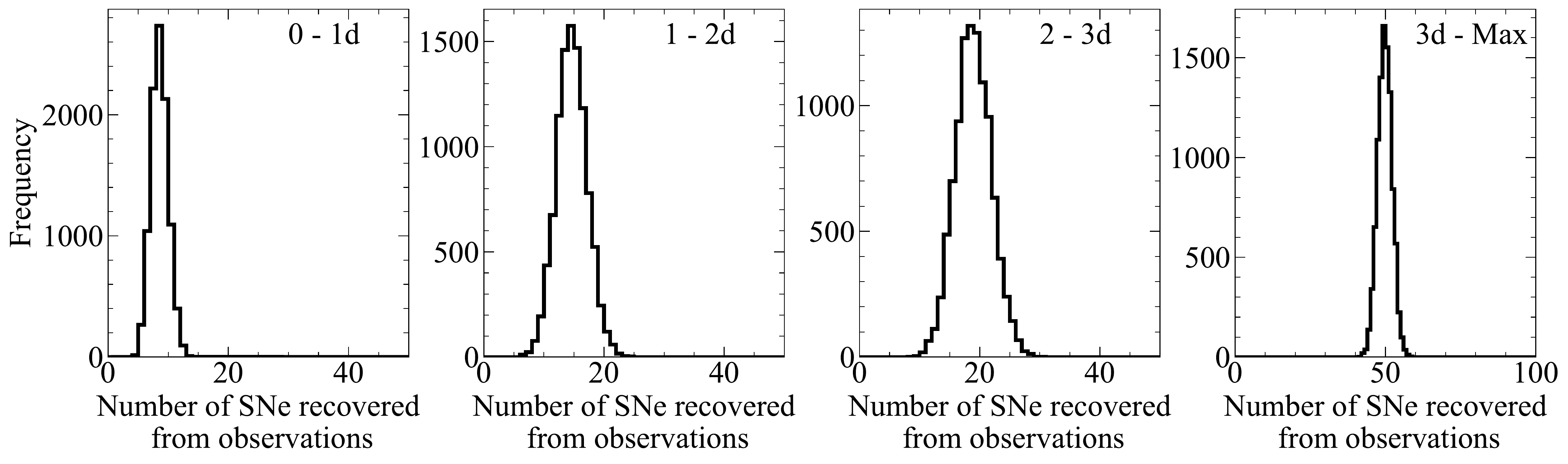}
\caption{ Number of SNe~Ia recovered from ZTF within various time bins. The number of SNe~Ia recovered in each bin is sensitive to the assumed explosion date, which is not known precisely. We perform 10\,000 iterations whereby the explosion time of each SN is randomly sampled from the distributions found by \citealt{deckers--22}. Each histogram shows the spread in the number of SNe recovered assuming these explosion times.}
\label{fig:obs_dist}
\centering
\end{figure*}

%

\subsection{The intrinsic fraction of SNe~Ia with an early excess}
\label{sect:excess_rate}

The agreement between the SNe~Ia rates calculated using our simulations in Sect.~\ref{sect:ia_rate} and literature values demonstrates the validity of these simulations. Given this, we now look to use \texttt{simsurvey} to constrain the fraction of SNe~Ia that could have an early excess. If a large fraction of SNe~Ia result from mechanisms that produce an excess, the number of SNe discovered shortly after explosion should be higher than if only a few SNe contain an excess. Therefore, given the sample of young SNe~Ia from \cite{yao--19}, $\mathcal{O}$, we wish to calculate the most likely fraction of SNe~Ia with an excess, $\mathcal{R}_x$, that reproduces the number of SNe discovered by ZTF in the sample. In other words the highest $P(\mathcal{R}_x | \mathcal{O})$. 

\par

To do this, we perform additional simulations whereby we vary the fraction of SNe~Ia with an excess and count the number, $\mathcal{N}$, that meet the selection criteria used in \cite{yao--19} and Sect.~\ref{sect:ia_rate} at various time intervals. As we are interested in directly comparing the number of SNe~Ia from our simulations to those observed, we use the rate calculated in Sect.~\ref{sect:ia_rate} ($2.56\times 10^{-5}$~SNe~yr$^{-1}$~Mpc$^{-3}$) rather than simply injecting a fixed number of 100\,000 transients. We assume that $\mathcal{R}_x$ is between 0 -- 100\% and repeat our simulations 10\,000 times with the injected models randomly sampled for each iteration from those presented in Sect.~\ref{sect:model_lcs}. We apply the same cuts to our simulations as used in \cite{yao--19} and Sect.~\ref{sect:ia_rate}. We further count the number of SNe passing these cuts that were observed between 0 -- 1\,d, 1 -- 2\,d, and 2 -- 3\,d after explosion, and between 3\,d after explosion and maximum light. In focusing on the very early times in this way, we will be more sensitive to the number of SNe~Ia with an excess. Figure~\ref{fig:counts_dist} shows the distributions of the number of SNe~Ia recovered across the 10\,000 iterations for each $\mathcal{R}_x$. Unsurprisingly, Fig.~\ref{fig:counts_dist} shows that as the fraction of SNe containing an excess increases, the number of discoveries at early times also increases. For example, if 100\% of SNe~Ia contained an excess, our simulations indicate that a median of $\sim$22 would be detected earlier than 1\,d. In comparison, if 0\% of SNe~Ia contained an excess, only a median of $\sim$2 would be detected.

\par

\begin{figure}
\centering
\includegraphics[width=\columnwidth]{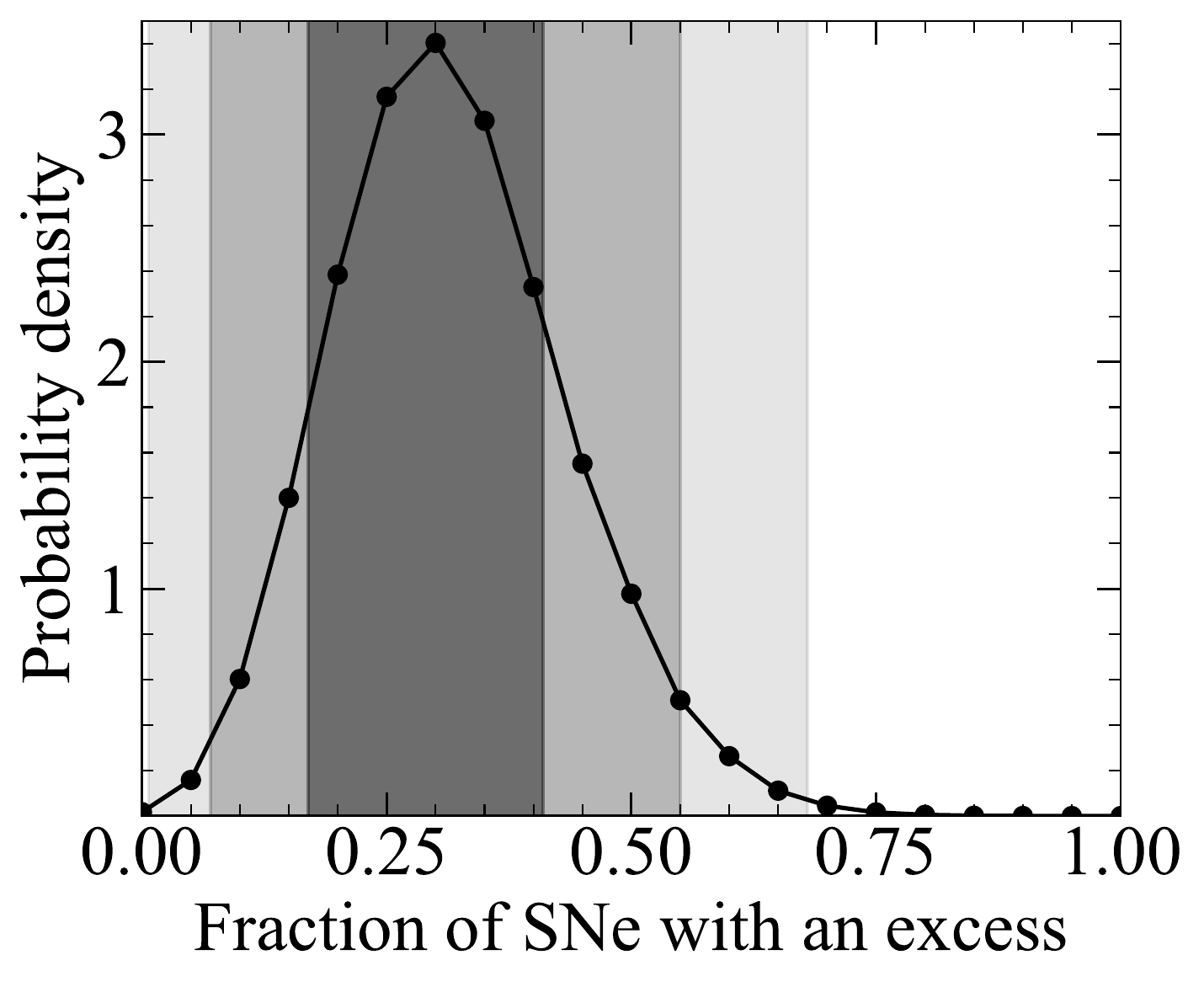}
\caption{Probability density distribution for the fraction of SNe~Ia containing an excess. Shaded regions show the 1, 2, and 3$\sigma$ range.
}
\label{fig:excess_prob_dist}
\centering
\end{figure}

The distributions shown in Fig.~\ref{fig:counts_dist} give the probability of detecting $\mathcal{N}$ SNe in each time bin, assuming some fraction, $\mathcal{R}_x$, of the intrinsic population contains an excess, $P(\mathcal{N} | \mathcal{R}_x)$. From Bayes theorem, $P(\mathcal{N} | \mathcal{R}_x) \propto P(\mathcal{R}_x | \mathcal{N})$. Hence, if $\mathcal{N}$ was exactly known from the observations, then to determine $P(\mathcal{R}_x | \mathcal{O})$ we would simply need to count the number of simulations, out of 10\,000, for which $\mathcal{N}$ SNe were detected. The explosion time of each SN however is not known precisely and therefore the number of SNe detected within a given bin will vary depending on the assumed explosion dates. To account for this we marginalise over the explosion times. By comparing to models from \cite{magee--20}, \cite{deckers--22} estimated the explosion time for most SNe~Ia in the \cite{yao--19} sample. For 10\,000 iterations, we randomly sample the probability distributions for the explosion time of each SN from \cite{deckers--22} and calculate the number of SNe recovered within the time bins. These distributions are shown in Fig.~\ref{fig:obs_dist} and hence provide the probability of detecting $\mathcal{N}$ SNe given the sample of observations $\mathcal{O}$, $P(\mathcal{N} | \mathcal{O})$. The probability $P(\mathcal{R}_x | \mathcal{O})$ is therefore given by the summation over all possible values of $\mathcal{N}$ and the product of individual probabilities within each time bin, $j$
\begin{equation}
P(\mathcal{R}_x | \mathcal{O}) \propto \prod_{j} \left( \sum_{\mathcal{N} = 0}^\infty P(\mathcal{N} | \mathcal{O}) P(\mathcal{N} | \mathcal{R}_x) \right)_j.
\end{equation}
The resultant probability distribution for the fraction of SNe~Ia containing an excess is shown in Fig.~\ref{fig:excess_prob_dist}.

\par

From Fig.~\ref{fig:excess_prob_dist}, we find $\mathcal{R}_x \sim 28^{+13}_{-11}\%$. With $\gtrsim$95\% confidence, we can exclude that $\lesssim$7\% or $\gtrsim$55\% of SNe~Ia contain an excess. Therefore, while some SNe~Ia show signs of an excess, it is highly likely that most SNe~Ia do not result from mechanisms that produce early excess emission. In addition, they could be somewhat common and account for approximately one in four  SNe~Ia.

%

\section{Discussion}
\label{sect:discuss}

\subsection{Comparisons with previous estimates of the early excess rate}

Comparing the recovered SNe~Ia from our simulations to those recovered within ZTF, we estimate that the intrinsic rate of SNe~Ia with an early excess is $\sim28^{+13}_{-11}\%$ of the overall population. We also exclude fractions of $\lesssim$7\% or $\gtrsim$55\% at a 95\% confidence level (Fig.~\ref{fig:excess_prob_dist}).

\par

Based on various samples of young SNe~Ia, the fraction observed with an excess of flux at early times is estimated to range from 14 -- 35\%. \cite{magee--20} present a series of radiative transfer models assuming different \nifs{} distributions and demonstrate the strong impact this has on the luminosity at early times. Comparing these models to a sample of literature objects observed approximately two weeks before maximum light, they estimate that $\sim$22\% of the SNe~Ia in their sample are inconsistent with their models and hence could show evidence for early excesses. \cite{deckers--22} use the same method of model comparisons and compare to the sample of young ZTF SNe~Ia from \cite{yao--19} used in Sect.~\ref{sect:rates} for our rates calculations. They argue that $\sim$18\% could show signs of an excess.

\par

Rather than comparisons against model light curves, \cite{jiang--18} use a visual classification scheme to estimate the early excess fraction. Here, early excesses are defined as either `spikes', `bumps', or generally having a broad light curve (relative to SN~2012cg). \cite{jiang--18} estimate that $\sim$35\% of the SNe in their gold sample can be classified as having an early excess. They also note however that the fraction of SNe~Ia with an excess is highly dependent on the SN subclass. Finally, \cite{jiang--20} present a handful of young SNe~Ia observed as part of the HSC-SSP survey and estimate an early excess rate of $\sim$14\% based on power-law fits to the early light curve.

\par

The previously mentioned studies estimate the fraction of SNe~Ia observed with an excess. As they are based on which SNe are observed, this may not be reflective of the intrinsic excess population. For example, SNe~Ia with an excess are brighter at early times and hence could be over-represented in such samples. At the same time, if the explosion epoch is estimated to be later than the true value then the lack of an excess could simply be due to the fact that it has faded away by the beginning of observations and not that no excess was present. These cases would cause the excess fraction to be under-estimated. In contrast, our simulations and analysis naturally account for excess SNe~Ia being brighter and uncertainty in the explosion epochs. We find that our estimate for the intrinsic rate of $\sim28^{+13}_{-11}\%$ is consistent with most estimates for the observed rate from the literature.

\par

\subsection{Implications for SNe~Ia progenitors and explosion mechanisms}

From our simulations of multiple different scenarios producing flux excesses at early times (such as double detonations, \nifs{} clumps, and companion or CSM interaction), we find that overall such excesses are somewhat rare, but could account for approximately one in four SNe~Ia. In spite of this, many of the individual mechanisms proposed struggle in some way with reproducing the majority of the flux excess SNe~Ia that have been observed to date.

\par

\cite{kasen--10} estimate that, for all types of companion stars, the signature due to companion interaction would only be observable in $\lesssim$10\% of cases. The observed number of young SNe~Ia from \cite{yao--19} is consistent with a population in which 10\% contain an excess (within the 95\% confidence region), however the companion interaction scenario struggles to reproduce many of the observed flux excesses. For example, the best fitting model presented by \cite{hosseinzadeh--17} for the early light curve of SN~2017cbv over-predicts the UV flux. A similar issue is also observed by \cite{miller--19yvq} fitting the optical colours of SN~2019yvq. The disagreement between early light curves of this scenario and observations could result from some assumptions or missing physics, however companion interaction is also expected to produce late-time H or He emission, which has yet to be observed in early excess SNe~Ia (e.g. \citealt{sand--18, siebert--20}). 

\par

Many of the observable signatures of companion interaction are similar to those produced by interaction with CSM, which has also been suggested for some SNe~Ia with flux excesses. Here, the lack of strong H or He emission at late times could be explained if CSM was H- and He-deficient and instead dominated by other elements, such as C. This may be similar to what is required for some SNe~Ia with excesses, however it also brings challenges if one attempts to explain the majority of the population. For example, \cite{jiang--18} argue that 100\% of the 91T-like SNe in their sample show evidence for early excesses. They argue that the short-lived C features seen in such SNe are inconsistent with a C/O-dominated CSM such as that produced by a merger. More models of interaction with H- and He-deficient CSM are required to fully understand the signatures that should be expected and hence the extent to which this scenario may or may not be excluded for some or all early excess SNe~Ia. 

\par

Similarly, mechanisms producing red excesses, such as \nifs{} clumps or double detonation explosions, are also likely ruled out as producing a significant fraction of SNe~Ia with excesses. In these cases, while the decay of radioactive material near the surface of the ejecta can reproduce the shape of the excess, the colours produced around maximum light are significantly redder than normal SNe~Ia, and indeed most early excess SNe~Ia (e.g. \citealt{magee--20b, magee--21}). Some red early excess SNe~Ia do exist however, such as SN~2016jhr, which \cite{jiang--2017} argue is consistent with a double detonation explosion.

\par

In summary, considering each of the different early excess scenarios used as part of this work, many of them produce similar observables (in terms of the excess luminosity and duration) and hence may be considered equally viable from our rate estimate. More detailed studies including colour and spectral evolution however have shown that this is not the case and therefore it is unlikely that any one of the scenarios considered here could fully explain the $\sim28^{+13}_{-11}\%$ of SNe~Ia with an excess that is estimated from our simulations. Each of the scenarios may be relatively rare on their own, but given their similar light curve shapes at early times the combined effect overall could be that early excesses in general are more common.

%

\section{Conclusions}
\label{sect:conclusions}

In this work, we presented the first simulations specifically testing the recovery efficiency of young SNe~Ia with and without an excess of flux. Such excesses have only been observed in a handful of cases, however it is unclear if this is because they are intrinsically rare or if this is due to the difficulty in observing SNe~Ia sufficiently early.

\par

Using \texttt{simsurvey}, we simulated four survey strategies, each based on observing logs from ZTF. As input for our simulations, we used existing radiative transfer models of SNe~Ia from the literature. This set of models covers a broad range of early light curve luminosities and shapes, and includes models producing early excesses from companion interaction, \nifs{} clumps, and helium shell detonations.

\par

For each survey strategy, we ran multiple simulations with each of the different input models considered. Our simulations show that the typical recovery efficiency of the ZTF 3\,d cadence public survey (where recovery is defined as having two 5$\sigma$ detections) is $\sim$52\%. In this survey, we also find that only $\lesssim$1.5\% of detections occur within the first 3\,d following explosion for models without an excess. Models with an excess are of course more luminous at early times and therefore $\sim$5.6 -- 7.3\% of detections occur within the first 3\,d. Since late 2020, the ZTF public survey has increased its observing cadence to 2\,d and our simulations indicate that this should result in a $\sim$50\% increase in the number of SNe~Ia detections made during the first few days of explosion. Our simulations also show however, that if all SNe~Ia did contain an excess, $\lesssim$30\% would be detected early enough for this to be observed, even for a 2\,d cadence survey. 

\par

From our survey simulations, we determine the recovery efficiency as a function of absolute magnitude and redshift. Using these recovery efficiencies we calculate the rate of SNe~Ia within ZTF by applying the same selection criteria as \cite{yao--19} and comparing to their sample of young SNe~Ia. Using this SNe~Ia rate, we perform additional simulations whereby we vary the fraction of SNe~Ia that contain an excess. For each fraction, we calculate the number of young SNe~Ia expected to be discovered within various time bins. By comparing to the number of SNe~Ia in these same time bins from the \cite{yao--19} sample, we find that $\sim28^{+13}_{-11}\%$ of SNe~Ia could contain an excess, which is consistent with previous estimates from the literature.

\par

The origin of early excesses in SNe~Ia is unknown. Multiple scenarios have been proposed and, given the diversity in excesses observed, it is probable that multiple scenarios are required. While our simulations have constrained the excess rate to $\sim28^{+13}_{-11}\%$, to place limits on the rate of each scenario is a challenging prospect and likely requires a larger sample of observed SNe~Ia with excesses. Fortunately, the increased cadence of ZTF~II will provide a significant increase in the number of young SNe~Ia detected, which will prove invaluable for adding further constraints to the early excess origins.

%

\section*{Acknowledgements}

We thank the referee for their constructive comments, which helped to focus our manuscript. We thank T. Collett for useful discussions. MRM and KM are funded by the EU H2020 ERC grant no. 758638. This work has received funding from the European Research Council (ERC) under the European Union’s Horizon 2020 research and innovation programme (LensEra: grant agreement No 945536). Based on observations obtained with the Samuel Oschin Telescope 48-inch and the 60-inch Telescope at the Palomar Observatory as part of the Zwicky Transient Facility project. Major funding has been provided by the U.S National Science Foundation under Grant No. AST-1440341 and by the ZTF partner institutions: the California Institute of Technology, the Oskar Klein Centre, the Weizmann Institute of Science, the University of Maryland, the University of Washington, Deutsches Elektronen-Synchrotron, the University of Wisconsin-Milwaukee, and the TANGO Program of the University System of Taiwan. M.~W.~C acknowledges support from the National Science Foundation with grant number PHY-2010970
\section*{Data Availability}
All models presented in this work are available on GitHub\footnote{\href{https://github.com/MarkMageeAstro/TURTLS-Light-curves}{https://github.com/MarkMageeAstro/TURTLS-Light-curves}}.



\bibliographystyle{mnras}
\bibliography{Magee}




\appendix
\section{Additional plots}

\begin{figure*}
    \centering
    \begin{subfigure}[b]{\textwidth}
        \includegraphics[width=\textwidth]{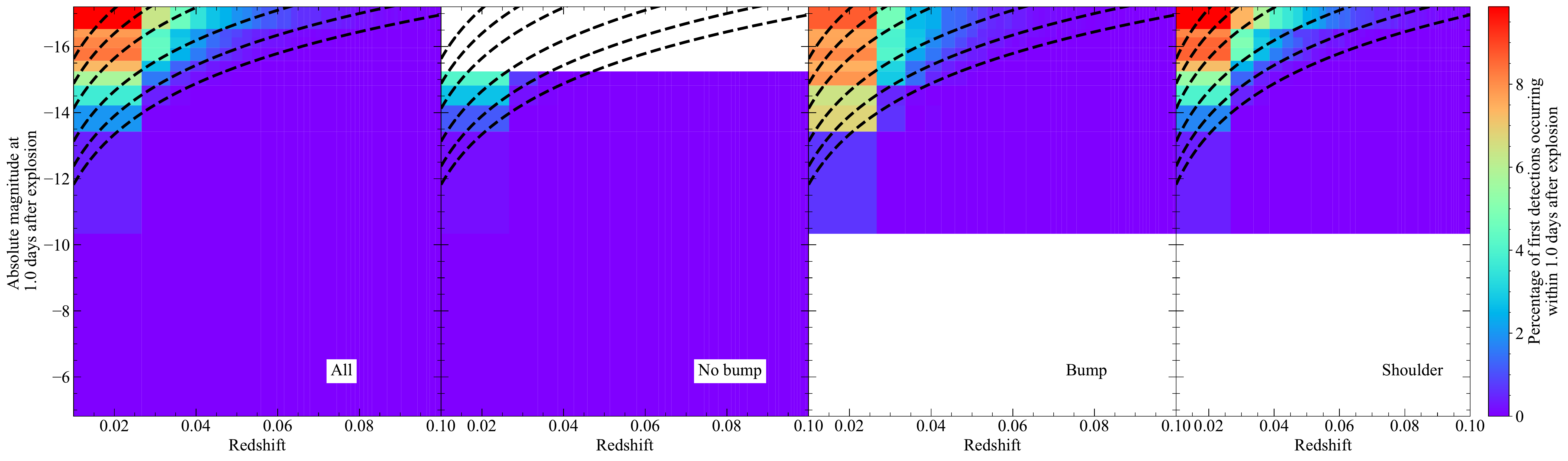}
    \end{subfigure}
    \\
    \begin{subfigure}[b]{\textwidth}
        \includegraphics[width=\textwidth]{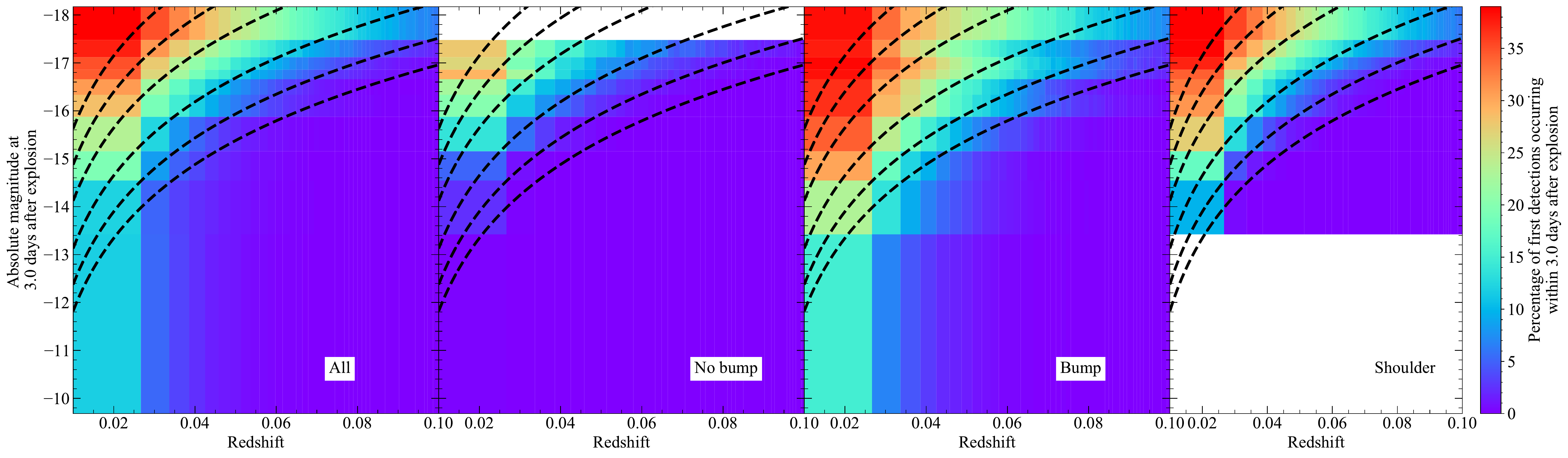}
    \end{subfigure}

    \caption{As in Fig.~\ref{fig:y1_mag_v_z} for the ZTF~I public survey plan.
   }
    \label{fig:y1_pub_mag_v_z}
\end{figure*}

\begin{figure*}
    \centering
    \begin{subfigure}[b]{\textwidth}
        \includegraphics[width=\textwidth]{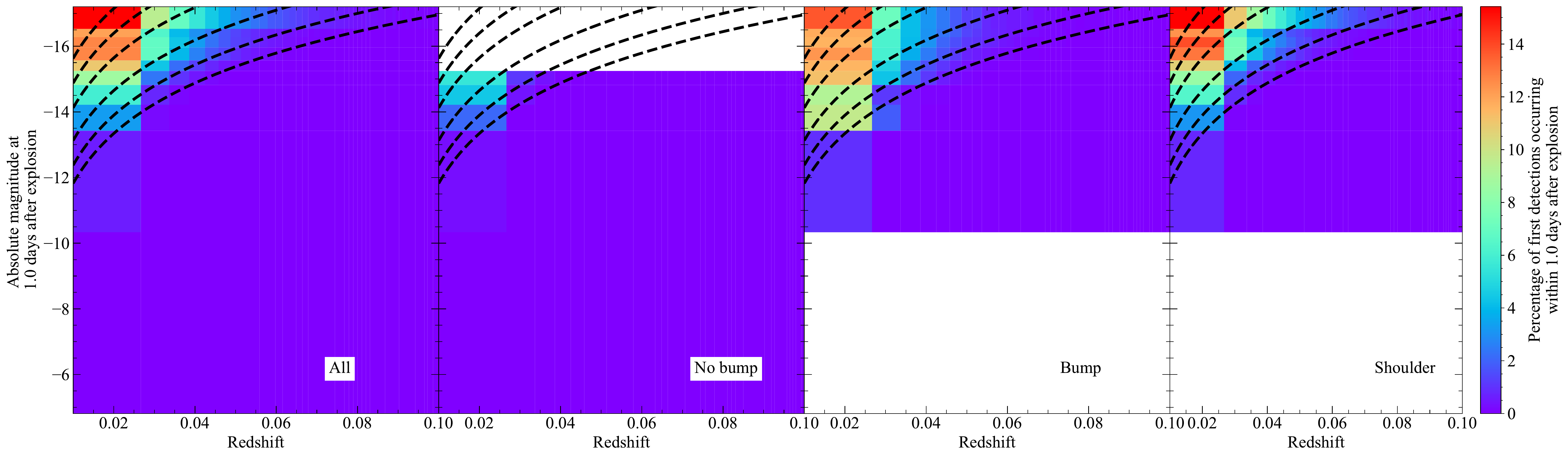}
    \end{subfigure}
    \\
    \begin{subfigure}[b]{\textwidth}
        \includegraphics[width=\textwidth]{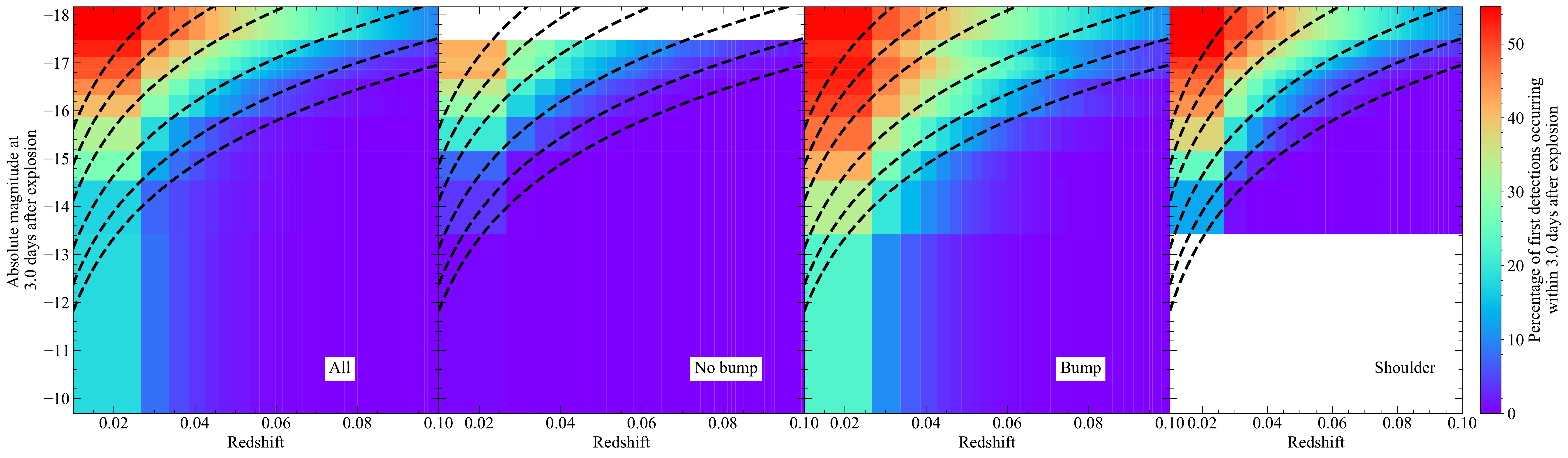}
    \end{subfigure}

    \caption{As in Fig.~\ref{fig:y1_mag_v_z} for the ZTF~II public survey plan.
   }
    \label{fig:ztfii_pub_mag_v_z}
\end{figure*}

\begin{figure*}
    \centering
    \begin{subfigure}[b]{\textwidth}
        \includegraphics[width=\textwidth]{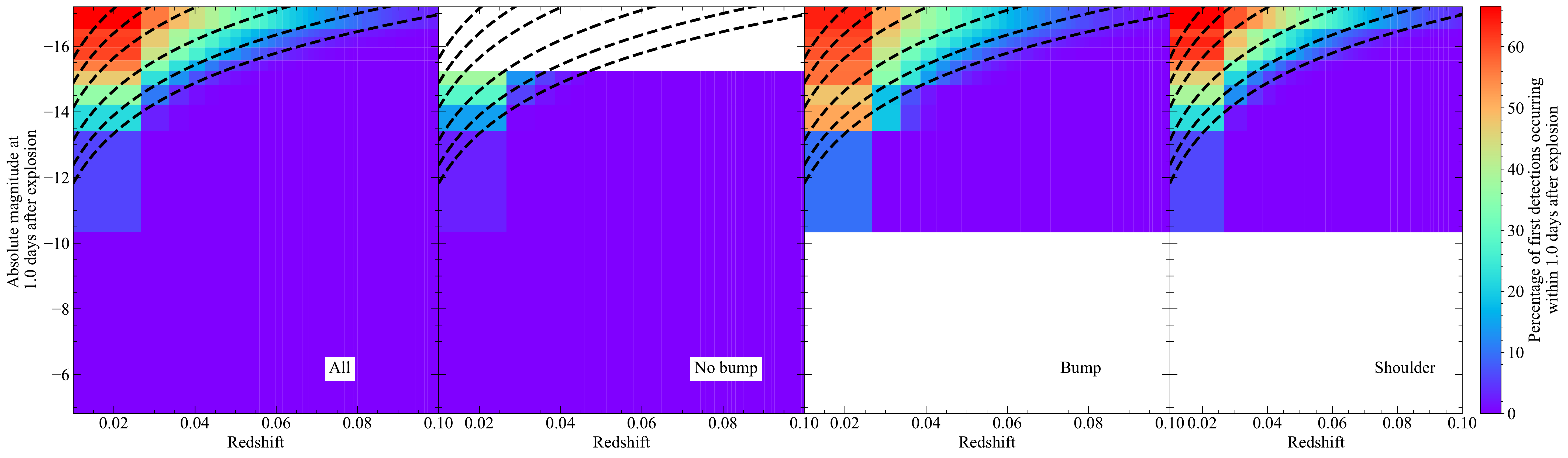}
    \end{subfigure}
    \\
    \begin{subfigure}[b]{\textwidth}
        \includegraphics[width=\textwidth]{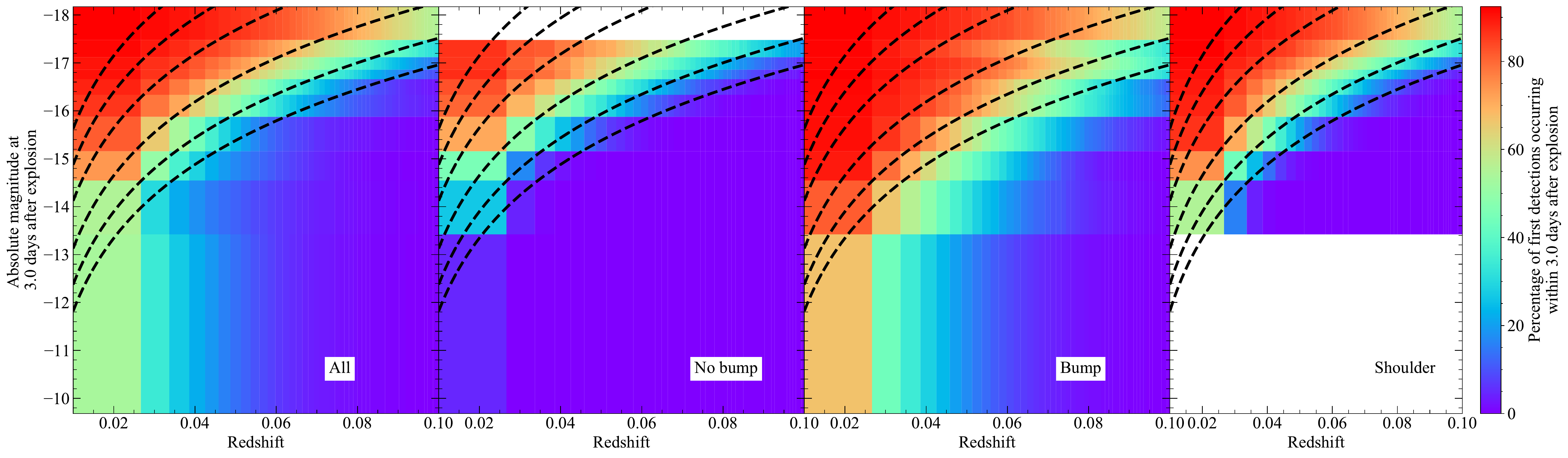}
    \end{subfigure}

    \caption{As in Fig.~\ref{fig:y1_mag_v_z} for the high cadence survey plan.
   }
    \label{fig:hc_mag_v_z}
\end{figure*}



\bsp	
\label{lastpage}
\end{document}